% Macros and definitions for manuscript to be mailed to J. Phys. A.
%
% Define page size, baselineskip for double spacing
%
\magnification=\magstep1
\headline={\ifnum\pageno=1\hfil\else\hfil\tenrm--\ \folio\ --\hfil\fi}
\footline={\hfil}
\hsize=6.0truein
\vsize=8.54truein
\hoffset=0.25truein
\voffset=0.25truein
\baselineskip=13pt
%
% No hyphenation
%
\tolerance=9000
\hyphenpenalty=10000
%
% FONTS
%
% smcap is small capitals for lower case characters.
% Used in authors name, postal addr., and table headings if desired.
%

% a boldface math font
\font\mbf=cmmib10 \font\mbfs=cmmib10 scaled 833
% a boldface math symbol font
\font\msybf=cmbsy10 \font\msybfs=cmbsy10 scaled 833
\font\smcap=cmcsc10
%
% MATH MACROS
%
% Using exercise 17.20 to define the bold face math symbols
% Others can be added as needed.
% Use in this way ${\bmsy\grad\cdot}{\bmit\xi} =0$. "div \xi=0"
%

\textfont9=\mbf \scriptfont9=\mbfs \scriptscriptfont9=\mbfs

\textfont10=\msybf \scriptfont10=\msybfs \scriptscriptfont10=\msybfs
%
% Define Greek letters to allow use when \mbf is active font
%
\mathchardef\alpha="710B
\mathchardef\beta="710C
\mathchardef\gamma="710D
\mathchardef\delta="710E
\mathchardef\epsilon="710F
\mathchardef\zeta="7110
\mathchardef\eta="7111
\mathchardef\theta="7112
\mathchardef\iota="7113
\mathchardef\kappa="7114
\mathchardef\lambda="7115
\mathchardef\mu="7116
\mathchardef\nu="7117
\mathchardef\xi="7118
\mathchardef\pi="7119
\mathchardef\rho="711A
\mathchardef\sigma="711B
\mathchardef\tau="711C
\mathchardef\upsilon="711D
\mathchardef\phi="711E
\mathchardef\chi="711F
\mathchardef\psi="7120
\mathchardef\omega="7121
\mathchardef\varepsilon="7122
\mathchardef\vartheta="7123
\mathchardef\varpi="7124
\mathchardef\varrho="7125
\mathchardef\varsigma="7126
\mathchardef\varphi="7127
%define \nabla to allow use when \msybf is active font
\mathchardef\nabla="7272
%define \cdot to allow use when \msybf is active font
\mathchardef\cdot="7201
 % rename \nabla as \grad
%
% \lta and \gta produce > and < signs with twiddle underneath
% (from S. Tremaine)
%
\def\spose#1{\hbox to 0pt{#1\hss}}
\def\lta{\mathrel{\spose{\lower 3pt\hbox{$\mathchar"218$}}
     \raise 2.0pt\hbox{$\mathchar"13C$}}}
\def\gta{\mathrel{\spose{\lower 3pt\hbox{$\mathchar"218$}}
     \raise 2.0pt\hbox{$\mathchar"13E$}}}
%
% \ldb and \rdb are double brackets, for use in formulae
%

%
% \bigldb and \bigrdb are \big size double brackets
%

%
% \biggldb and \biggrdb are \bigg size double brackets
%

%
% UNITS MACROS
%
% Roman lettered units for use in math mode
%

%
% \arcsec produces arcsec symbol so that 3\arcsec5 produces 3."5 with the
% second symbol and the period aligned.
%

%
% EQUATION NUMBERING MACROS
%
\newcount\eqnumber
\eqnumber=1
% \new macro produces sequentially numbered equations by writing \eqno(\new)
% at end of displayed equations
%
\def\new{{\the\eqnumber}\global\advance\eqnumber by 1}
%
% To refer to an equation which is 5 equations back, write "equation (\ref5)"
%
\def\ref#1{\advance\eqnumber by -#1 {\the\eqnumber}\advance
     \eqnumber by #1}
%
% \last macro is like \new except counter is not advanced. Useful for
% equations which are in parts a and b.
%
\def\last{\advance\eqnumber by -1 {\the\eqnumber}\advance 
     \eqnumber by 1}
%
% To name an equation, place command "\eqnam{\Poisson}" before equation, and
% thereafter "equation(\Poisson)" will generate the proper equation number.
%
\def\eqnam#1{\xdef#1{\the\eqnumber}}
%
% REFERENCE MACROS
%
% To generate reference to a paper in Ap.J. volume 300, p.123 
% write \apj{Claus, S. 1990}{300}{123}
%
\def\refindent{\par\noindent\hangindent=3pc\hangafter=1 }

\def\apj#1#2#3#4{\refindent{#1}\ #2\ {\it Astrophys.\ J.}\ {\bf#3} #4}

\def\icarus#1#2#3#4{\refindent{#1}\ #2\ {\it Icarus} {\bf#3} #4}

\def\refpaper#1#2#3#4#5{\refindent{#1}\ #2\ {\it #3}\ {\bf#4} #5}
\def\refbook#1#2#3{\refindent{#1}\ #2\ #3}
%
% \refrule produces a horizontal rule used in references with identical
% authors.
%
\def\refrule{\hbox to 3pc{\leaders\hrule depth-2pt height 2.4pt\hfill}}
\def\etal{{\it et al}}
%
%
% Define macros to start sections or subsections of the paper
% e.g.  \sect{1.~INTRODUCTION}
% e.g.  \subsec{2.2}{Mode Classification}
% N.B.  I don't use \subsec for subsection 1 if it comes after the
% section title as there is too much space. use 
% \centerline{2.1.~{\it Equilibrium Models}}
%
\def\sect#1 {
  \bigbreak
  \centerline{\bf #1}
  \bigskip}
\def\subsec#1#2 {
  \bigbreak
  \centerline{#1.~{\it #2}}
  \bigskip}
%
% That's all folks!

\def\tcrit{t_{\rm crit}}
\def\mtil{{\tilde m}}
\def\Nbin{N_{\rm bin}}
\def\Ndec{N_{\rm dec}}
\def\Nbd{N_{bd}}
\def\Nmax{N_{\rm max}}
\def\nmin{n_{\rm min}}
\def\mmax{m_{\rm max}}
\def\dlnndlnm{d \ln n/d \ln m}
\def\dlnpdlnx{d \ln \varphi/d \ln x}

\centerline{\bf A SURVEY OF NUMERICAL SOLUTIONS TO THE}
\centerline{\bf COAGULATION EQUATION}

\bigskip
\centerline{\smcap Man Hoi Lee}
\bigskip
\centerline{\it Department of Physics, University of California,
                Santa Barbara, CA 93106, USA}
\bigskip
\centerline{E-mail: mhlee@europa.physics.ucsb.edu}

% \vfill

% \noindent{Running Head: NUMERICAL SOLUTIONS TO COAGULATION EQUATION}

% \eject

\sect{ABSTRACT}

We present the results of a systematic survey of numerical solutions to the
coagulation equation for a rate coefficient of the form $A_{ij} \propto
(i^\mu j^\nu + i^\nu j^\mu)$ and monodisperse initial conditions.
The results confirm that there are three classes of rate coefficients with
qualitatively different solutions.
For $\nu \le 1$ and $\lambda = \mu + \nu \le 1$, the numerical solution
evolves in an orderly fashion and tends toward a self-similar solution at
large time $t$.
The properties of the numerical solution in the scaling limit agree with
the analytic predictions of van Dongen and Ernst.
In particular, for the subset with $\mu > 0$ and $\lambda < 1$, we disagree
with Krivitsky and find that the scaling function approaches the
analytically predicted power-law behavior at small mass, but in a damped
oscillatory fashion that was not known previously.
For $\nu \le 1$ and $\lambda > 1$, the numerical solution tends toward a
self-similar solution as $t$ approaches a finite time $t_0$.
The mass spectrum $n_k$
develops at $t_0$ a power-law tail $n_k \propto k^{-\tau}$ at large mass
that violates mass conservation, and runaway growth/gelation is expected to
start at $\tcrit = t_0$ in the limit the initial number of particles
$n_0 \to \infty$.
The exponent $\tau$ is in general less than the analytic prediction
$(\lambda + 3)/2$, and $t_0 = K/[(\lambda - 1)n_0 A_{11}]$ with
$K = 1$--$2$ if $\lambda \gta 1.1$.
For $\nu > 1$, the behaviors of the numerical solution are similar to those
found in a previous paper by us.
They strongly suggest that there are no self-consistent solutions at any
time and that runaway growth is instantaneous in the limit $n_0 \to \infty$.
They also indicate that the time $\tcrit$ for the onset of runaway growth
decreases slowly toward zero with increasing $n_0$.

\vfill\eject

\sect{1.~INTRODUCTION}

Smoluchowski's coagulation equation is the mean-rate equation that
describes the evolution of the mass spectrum of a collection of particles
due to successive mergers.
It is widely used for modeling growth in many fields of science.
Examples include planetesimal accumulation, mergers in dense clusters of
stars, coalescence of interstellar dust grains, and galaxy mergers in
astrophysics, aerosol coalescence in atmospheric physics, colloids, and
polymerization and gelation (see, e.g., Drake 1972, Ernst 1986, Jullien and
Botet 1987, Lee 1993, 2000, and references therein).

If the masses of the particles are integral multiples of a minimum mass
$m_1$, the coagulation equation is written in discrete form as
$$
{dn_k \over dt} = {1 \over 2} \sum_{i+j = k} A_{ij} n_i n_j -
                  n_k \sum_{i=1}^\infty A_{ki} n_i
\eqnam{\dCoEq}
\eqno(\new)
$$
where $n_k$ is the number of particles of mass $m_k = k m_1$ in a volume
$V$ and $A_{ij}$ is the rate coefficient (or coagulation kernel) for mergers
between particles of mass $m_i$ and $m_j$.\footnote{$^1$}
{We can interpret $n_k$ as the concentration (i.e., the number of
particles per unit volume) if we replace $A_{ij}$ by $A'_{ij} = V A_{ij}$.}
In equation (\dCoEq), it is assumed that the merging of two particles of mass
$m_i$ and $m_j$ results in one particle of mass $m_i + m_j$.
The coagulation equation can also be written in continuous form as
$$
\eqalign{
{dn(m) \over dt} &= {1 \over 2} \int_0^m dm' \, A_{m',m-m'} n(m') n(m-m')\cr
     &\qquad\qquad\qquad\qquad - n(m) \int_0^\infty dm' \, A_{m,m'} n(m')
}
\eqnam{\cCoEq}
\eqno(\new)
$$
where $n(m) dm$ is the number of particles of mass between $m$ and $m+dm$
and $A_{m,m'}$ is the rate coefficient for mergers between particles of mass
$m$ and $m'$.

Examples of the rate coefficient $A_{ij}$ as a function of $m_i$ and $m_j$
(or equivalently $i$ and $j$) that arise in various problems can be
found in the references cited above.
Most rate coefficients used in the literature are homogeneous
functions of degree $\lambda$, i.e., $A_{ai,aj} = a^\lambda A_{ij}$.
The exponent $\lambda$ specifies the mass dependence of the probability of
merger for two particles of comparable mass ($i \sim j$).
It is also useful to classify $A_{ij}$ according to the exponents $\mu$ and
$\nu$ for the merger between a light particle and a heavy particle:
$A_{ij} \propto i^\mu j^\nu$ for $i \ll j$ and $ \mu + \nu = \lambda$
(see, e.g., Ernst 1986).
For example, $A_{ij} \propto i+j$ has $\mu=0$, $\nu=1$, and $\lambda=1$.

For a few simple rate coefficients and monodisperse initial
conditions (i.e., $n_0$ particles of mass $m_1$ at $t=0$), there are exact
analytic solutions to the discrete form of the coagulation equation
(Trubnikov 1971, Ernst 1986, and references therein).
The analytic solutions for $A_{ij} \propto$ constant and $i+j$ show orderly
evolution of a smooth mass spectrum at all times, and they agree with the
results from Monte Carlo simulations of the merger process in the limit
$n_0 \to \infty$ (with $n_0/V$ fixed).
These two cases are examples of orderly growth.
The analytic solution for $A_{ij} \propto ij$ develops a power-law tail
with $n_k \propto k^{-5/2}$ at large $k$ as $t \uparrow t_0 = 1/(n_0 A_{11})$.
This power-law tail violates mass conservation because it implies a
nonzero mass flux to the infinite-mass bin.
In this case, the results from Monte Carlo simulations in the limit
$n_0 \to \infty$ agree with the solution to the coagulation equation at
$t \le t_0$, but they show a transition from a smooth mass spectrum to a
smooth spectrum plus a massive runaway particle at $t = t_0$ (Spouge 1985,
Wetherill 1990).
The runaway particle (gel) acquires a mass much larger than that of the
other particles (sol) in the system and becomes detached from the smooth
mass spectrum of the rest of the particles at $t > t_0$.
This phenomenon is known as runaway growth in the astrophysics literature,
and the transition is considered to be the gelation transition in studies
of polymerization and gelation.

For most rate coefficients, there are no exact analytic solutions to
the coagulation equation.
However, there are extensive analytic results on the asymptotic properties
of the solutions for $A_{ij}$ with $\nu \le 1$ (see, e.g., review by Ernst
1986).
It is important to note that some of these analytic results (such as the
shape of the mass spectrum at small and large mass) are derived based
on assumptions (such as self-similar evolution) that have not been verified.
Nevertheless, the analytic results indicate that there are qualitatively
two types of solutions to the coagulation equation if $\nu \le 1$:

\item{(1)} if $\nu \le 1$ and $\lambda \le 1$, the solution shows orderly
growth at all times;
\item{(2)} if $\nu \le 1$ and $\lambda > 1$, the solution develops in a
finite time $t_0$ a power-law tail at large mass that violates mass
conservation.

\noindent For $A_{ij}$ with essentially $\nu < 1$, it has been
proved that a solution to the coagulation equation exists for all
times (including $t > t_0$ if $\lambda > 1$) and that the coagulation
equation is the limit of finite system (whether or not the runaway
particle and the other particles are allowed to interact at $t > t_0$ if
$\lambda > 1$)
(Leyvraz and Tschudi 1981, Spouge 1985, Bak and Heilmann 1994, Jeon 1998).
(For $\nu = 1$ and $\lambda > 1$, we have the example $A_{ij} \propto ij$,
where the coagulation equation needs to be modified for $t > t_0$ if there
is sol-gel interaction; Ziff \etal\ 1983, Bak and Heilmann 1994.)

Several authors have investigated the properties of the solutions
to the coagulation equation for $A_{ij}$ with $\nu > 1$, using series
expansion of the mass spectrum $n_k(t)$ about $t = 0$ and moments of
the mass spectrum (McLeod 1962, Hendriks \etal\ 1983, Ernst \etal\ 1984,
van Dongen 1987a).
The results suggest that

\item{(3)} if $\nu > 1$, there are no self-consistent solutions that
conserve mass at any time.

An alternative to the analytic approach is to solve the coagulation
equation numerically.
In Paper I (Lee 2000), a numerical code that can yield accurate solutions
to the discrete form of the coagulation equation, equation (\dCoEq), with a
reasonable number of numerical mass bins was developed.
The numerical code was used to study solutions to the coagulation equation
for $A_{ij}$ that are limiting cases for gravitational interaction.
We considered geometric or gravitational focusing dominated cross-section,
mass-independent or equipartition velocity dispersion, and the power-law
index of the mass-radius relation $\beta = 1/3$ (for planetesimals) or
$2/3$ (for stars).
For the two cases with geometric cross-section and $\beta = 1/3$, which
have $\nu \le 1$ and $\lambda \le 1$, the mass spectrum
evolves in an orderly fashion and tends towards a self-similar solution at
large time.
For the remaining cases, which have $\nu > 1$,
the numerical mass spectrum shows, after some evolution, an exponential
drop in an intermediate mass range and a power-law tail of the form
$n_k \propto k^{-\nu}$ (or $n(m) \propto m^{-\nu}$) at large mass.
This mass spectrum is not self-consistent because the power-law tail
implies a mass flux\footnote{$^2$}
{
As we pointed out in Paper I, in these cases, a massive particle grows
mainly by accumulating low-mass particles because of the much larger
number of low-mass particles.
So the growth rate of a massive particle is
${\dot m} = \int dm' A_{m,m'} n(m') m' \propto m^\nu$, since the integral
is dominated by the range $m' \ll m$.
Hence $n {\dot m}$ at the high-mass end of the mass spectrum is non-zero
and independent of $m$ if $n \propto m^{-\nu}$.
However, we did not point out that the mass flux from particles of mass
$m' \le m$ to particles of mass $m' > m$ is $F_m \approx m n(m) {\dot m}(m)
+ \int_m^{\mmax} dm' n(m') {\dot m}(m')$.
With $n {\dot m}$ independent of $m$, $F_m \approx n {\dot m} \mmax$,
which is independent of $m$ but increases with $\mmax$.
We have verified this by an explicit evaluation of the mass flux
(equation [12]) for the numerical solutions.
}
and, if $1 < \nu \le 2$, a cumulative mass that diverge with the maximum
particle mass, $\mmax$, included in the computational grid.
The time at which the power-law tail develops decreases toward
zero as the numerical parameter $\nmin$ decreases (see Section 3 for the
definition of $\nmin$).
Thus the numerical results strongly suggest that there are no
self-consistent solutions to the coagulation equation at any time if
$\nu > 1$.
We also considered a case with $\beta = 0$ as an example with
$\nu \le 1$ and $\lambda > 1$, and its mass spectrum develops a power-law
tail that violates mass conservation in a finite time $t_0$.
We discussed a simplified merger problem that illustrates the
qualitative differences in the solutions to the coagulation equation for
the three classes of $A_{ij}$.
The results in Paper I (and the analytic results cited above) strongly
suggest that there are two types of runaway growth.
For $A_{ij}$ with $\nu \le 1$ and $\lambda > 1$, runaway growth starts at a
finite time $\tcrit = t_0$, the time at which the coagulation equation
solution begins to violate mass conservation, in the limit $n_0 \to \infty$.
For $A_{ij}$ with $\nu > 1$, runaway growth is instantaneous
in the limit $n_0 \to \infty$, and there are indications (since decreasing
$\nmin$ is similar to increasing $n_0$) that the time $\tcrit$, in units of
$1/(n_0 A_{11})$, for the onset of runaway growth decreases slowly toward
zero with increasing initial number of particles $n_0$.
Recent Monte Carlo simulations have shown that the time for all particles to
coalesce into a single particle decreases as a power of the logarithm of
$n_0$ if $\nu > 1$ (Malyshkin and Goodman 2001; see also Spouge 1985,
Jeon 1999).

The study in Paper I was focused on the rate coefficient for
gravitational interaction, and the range of $\mu$ and $\nu$ studied was
limited.
In particular, $\mu$ was limited to $0$ and $\pm 1/2$, and the region
$\mu > 0$ and $\nu < 1$ was not studied at all.
Other authors have obtained numerical solutions to the coagulation equation
for rate coefficients that arise in specific problems.
However, we are not aware of any study that has systematically surveyed the
properties of numerical solutions as a function of $\mu$ and $\nu$ (and
$\lambda$) and compared them to the analytic results on the asymptotic
properties.
For such a study, it is important that the numerical code used can follow
the evolution of the mass spectrum accurately for a long time.
After the computation for this paper was nearly complete, it came to our
attention that Krivitsky (1995) has obtained numerical solutions to the
continuous form of the coagulation equation, equation (\cCoEq), for
$A_{m,m'} \propto (m m')^{\lambda/2}$ (which has $\mu = \nu = \lambda/2$)
and $A_{m,m'} \propto (m + m')^\nu$ (which has $\mu = 0$).
For $A_{m,m'} \propto (m m')^{\lambda/2}$ with $\lambda \le 1$, Krivitsky
found that the numerical solutions are self-similar at large time, but
that unlike the analytic result, the asymptotic behavior of the scaling
function at small mass is not a power law.
As we shall see, the latter result is incorrect because Krivitsky did not
evolve the numerical solutions for a sufficiently long time to see the true
asymptotic behavior at small mass.
The scaling function does in fact approach the analytically predicted
power law at small mass, but in a damped oscillatory fashion that was not
known previously.
It is unlikely that Krivitsky could have solved the coagulation equation
accurately for the necessary amount of time because the numerical code used
by Krivitsky does not conserve mass.
For $A_{m,m'} \propto (m + m')^\nu$ with $\nu > 1$, Krivitsky found that
the mass spectrum develops a slowly decreasing tail at very small time.
Only the numerical solution for $\nu = 2$ was shown.
Its evolution is qualitatively similar to that found in Paper I for
$\nu > 1$, but it is not clear that the tail is power-law in nature because
the maximum particle mass included in Krivitsky's computational grid is not
large enough.
It was also not demonstrated that the numerical solution is not sensitive
to the other numerical parameters.

In this paper we present the results of a systematic survey of numerical
solutions to the coagulation equation.
The purpose of this survey is
(1) to confirm that there are three classes of rate coefficients
with qualitatively different solutions to the coagulation equation and that
the boundaries of these three classes are as stated above;
(2) to investigate, in the cases where self-consistent solutions exist,
whether the solutions approach self-similar solutions as $t \to \infty$ or
$t \uparrow t_0$ and whether the scaling behaviors agree with the analytic
results;
(3) to study the dependence of $t_0$ on the exponents $\mu$, $\nu$, and
$\lambda$ for the runaway growth cases with $\nu \le 1$ and $\lambda >
1$; and
(4) to investigate whether the behaviors of the numerical solutions found
in Paper I for the cases with $\nu > 1$ are valid in general.
In Section 2 we describe the rate coefficient and the initial
conditions used in this survey.
In Section 3 we provide a brief summary of the numerical methods developed
in Paper I for solving the coagulation equation and additional information
on the accuracy of the numerical results.
The results are presented in Section 4, and the conclusions are summarized
in Section 5.

\sect{2.~RATE COEFFICIENT AND INITIAL CONDITIONS}

In this paper we consider a rate coefficient of the form
$$
A_{ij} = {1 \over 2} \left(i^\mu j^\nu + i^\nu j^\mu\right)
\eqnam{\Aij}
\eqno(\new)
$$
with $\mu \le \nu$.
Note that $A_{ai,aj} = a^{\mu + \nu} A_{ij}$ and
$A_{ij} \propto i^\mu j^\nu$ for $i \ll j$, consistent with the definitions
of the exponents $\mu$, $\nu$, and $\lambda = \mu + \nu$ in Section 1.
Since the rate coefficient in equation (\Aij) contains the exponents $\mu$ and
$\nu$ as parameters explicitly,
we can survey the entire $(\mu,\nu)$ space by varying $\mu$ and $\nu$.
This rate coefficient includes $A_{ij} = (ij)^{\lambda/2}$ (for $\mu =
\nu = \lambda/2$) and $A_{ij} = (i^\nu + j^\nu)/2$ (for $\mu = 0$), which
have been used to model polymerization (e.g., Hendriks \etal\ 1983), and
the cases $(\mu,\nu) = (1,4/3)$ and $(2,2)$, which have been used to model
planetesimal accumulation and stellar merger (Malyshkin and Goodman 2001).
It also has the nice property that it includes the three
cases with exact analytic solutions to the coagulation equation:
$A_{ij} = 1$, $(i+j)/2$, and $ij$ for $(\mu,\nu) = (0,0)$, $(0,1)$, and
$(1,1)$, respectively.

We have obtained numerical solutions to the discrete form of the
coagulation equation for the cases shown in figure 1.
The ranges of $\mu$ and $\nu$ considered contain most of the values
encountered in practical applications (but usually for other forms of
$A_{ij}$).
The numerical solutions were computed for the monodisperse initial
conditions with $n_0$ particles of mass $m_1$, i.e.,
$n_k(t=0) = n_0 \delta_{k1}$, where $\delta_{k1}$ is the Kronecker delta.
Hereafter, we adopt units such that $n_0 = 1$, $m_1 = 1$, and $A_{11} = 1$.
With this set of units, $m_k = k$ and time is in units of $1/(n_0 A_{11})$,
the timescale for every particle of mass $m_1$ to merge with another
particle of mass $m_1$.

\sect{3.~NUMERICAL METHODS}

In Paper I we have developed a numerical code that can yield accurate
solutions to the discrete form of the coagulation equation (equation [\dCoEq])
with a reasonable number of numerical mass bins.
A detailed description of the code can be found in Paper I.
In this section we provide a brief summary of the algorithm and its
numerical parameters.
We also provide additional information on the accuracy of the numerical
results.

Our numerical code uses a combination of linearly and logarithmically
spaced numerical mass bins.
The first $\Nbd$ numerical mass bins are linearly spaced with $\mtil_k = k$.
They have boundaries $\mtil_{k \pm 1/2} = k \pm 1/2$ and width
$\Delta \mtil_k \equiv \mtil_{k+1/2} - \mtil_{k-1/2} = 1$.
The next $\Nbd\times\Ndec$ numerical mass bins are logarithmically spaced,
with $\Nbd$ bins per decade of mass;
thus the $k$th mass is $\mtil_k = (\mtil_{k+1/2} + \mtil_{k-1/2})/2$, with
$\mtil_{k+1/2}/\mtil_{k-1/2} = 10^{1/\Nbd}$.
There are in total $\Nmax = (\Ndec+1)\Nbd$ mass bins, and the mass of the
most massive particles in the computational grid, $\mmax$, is
approximately $\Nbd 10^{\Ndec}$.
Initially, the number of ``active'' bins $\Nbin \ll \Nmax$.
At the end of each time step, $\Nbin$ is increased (if necessary) to include
all bins with $N_k > \nmin$, where $N_k$ is the total number of particles
in bin $k$ and $\nmin$ is a numerical parameter;
it is also increased if $N_{\Nbin+1}$ becomes comparable to the power-law
extrapolation from $N_{\Nbin}$.
Before $\Nbin$ reaches $\Nmax$, $\Nmax$ (or equivalently $\mmax$) has
no effects on the numerical results.
The numerical parameter $\nmin$ is specified in units of $n_0$ and, e.g.,
$\nmin = 10^{-30}$ in units of $n_0$ is equivalent to $\nmin = 1$ and
$n_0 = 10^{30}$ in physical units.
Thus the effect of the numerical parameter $\nmin$ is similar to not
allowing fractionally occupied numerical mass bins to interact.

The fundamental quantity evolved by our numerical code is the total mass
$M_k$ in bin $k$.
During a time step, the code calculates for each combination of $i$ and $j$
(with $i \le j \le \Nbin$) the mass loss from bins $i$ and $j$ due to
mergers between particles in those bins and distributes the total mass of
the merger products among the mass bins.
Thus the code conserves mass exactly.
For $i \le j \le \Nbd$, it is correct to assume that the merging particles
have masses $\mtil_i$ and $\mtil_j$ and that the merger products have mass
$\mtil_i + \mtil_j$.
For $i \le j$ and $j > \Nbd$, we assume that the particles in bin $i$ have
mass $\mtil_i$ (which is exact for $i \le \Nbd$) and that the mass
distribution within bin $j$ follows a power-law distribution:
$$
\rho_j(m) = c_j (m/\mtil_{j-1/2})^{q_j}
    \qquad {\rm for~} \mtil_{j-1/2} < m \le \mtil_{j+1/2}
\eqno(\new)
$$
where $\rho_j(m)dm$ is the total mass of particles with mass between $m$ and
$m + dm$.
The merger products have masses between $\mtil_i + \mtil_{j-1/2}$ and
$\mtil_i + \mtil_{j+1/2}$, and they are either added to a single bin $k$
(if $\mtil_{k-1/2} \le \mtil_i + \mtil_{j-1/2}$ and
$\mtil_i + \mtil_{j+1/2} \le \mtil_{k+1/2}$) or distributed between bins
$k$ and $k+1$.
In equation (\ref1), the power-law index $q_j$ is obtained from the masses in
the adjacent bins,
$$
q_j = {{\log\left({M_{j+1} \over \Delta \mtil_{j+1}} {\Big /}
                  {M_{j-1} \over \Delta \mtil_{j-1}}\right)} \over
       {\log\left(\mtil_{j+1}/\mtil_{j-1}\right)}}
\eqnam{\qj}
\eqno(\new)
$$
and the normalization constant $c_j$ from the constraint
$$
\int_{\mtil_{j-1/2}}^{\mtil_{j+1/2}} dm\,\rho_j(m) = M_j .
\eqno(\new)
$$

Our numerical code uses the second-order Runge-Kutta method with a variable
time step.
The time step is continuously adjusted so that the fractional change of each
$M_k$ per time step is less than $\delta_M$ and the mass loss from bin $k$
does not exceed $M_k$.

In Paper I we have compared in detail the numerical solutions from our
code to the exact analytic solutions for $A_{ij} = 1$, $(i+j)/2$, and $ij$,
with the last case at $t < t_0$ only.
The accuracy of the numerical solutions is extremely insensitive to
$\delta_M$ and $\nmin$ and improves rapidly with increasing $\Nbd$.
Hereafter, unless stated otherwise, the numerical results were obtained
using $\delta_M = 5\%$, $\nmin = 10^{-30}$ (in units of $n_0$), and
$\Nbd = 40$.

We report here several additional tests of our code.
Ziff (1980) has constructed three forms of rate coefficients, with a
parameter $\gamma$, for which a single moment ${\cal M}_\gamma(t) =
\sum_{k=1}^\infty m_k^\gamma n_k(t)$ of the mass spectrum $n_k(t)$ can be
calculated analytically.
We have obtained numerical solutions for a few of these rate coefficients
(including both orderly and runaway growth cases) and have confirmed that
the numerical results for the moment ${\cal M}_\gamma(t)$ agree with the
analytic results, with accuracy similar to what was found for the three
cases with exact analytic solutions.

In Section 4.3 we shall be interested in extending the calculations for
some of the runaway growth cases with $\nu \le 1$ and $\lambda > 1$ to
$t > t_0$.
Therefore, another test that we have performed is to extend the comparison
for the case $A_{ij} = ij$ to $t > t_0$.
For $A_{ij} = ij$, the evolution of the mass spectrum at $t > t_0$ depends
on whether or not the runaway particle (gel) and the other particles (sol)
interact.
The (unmodified) coagulation equation is valid if there is no sol-gel
interaction, and it has an exact analytic solution with $n_k(t) \propto
t^{-1} k^{-5/2}$ at large $k$ for all $t > t_0$ (Leyvraz and Tschudi 1981,
Ziff \etal\ 1983).
(The coagulation equation can be modified to take into account sol-gel
interaction, and an exact analytic solution also exists for this modified
coagulation equation; Ziff \etal\ 1983.)
Since our code does not take into account sol-gel interaction and does not
allow merger products with masses greater than $\mmax$ to interact,
we expect the numerical solution at $t > t_0$ to agree with the analytic
solution to the unmodified coagulation equation, except for $m_k \sim
\mmax$.
We have integrated the case $A_{ij} = ij$ up to $t = 1.25$ and have found
that the numerical solution at $t > t_0 = 1$ is in excellent agreement with
the analytic solution for $m_k \lta 0.01 \mmax$.

A quantity that will be discussed extensively in Section 4 is
the logarithmic slope $\dlnndlnm$ of the mass spectrum.
For the numerical results, we use
$$
\dlnndlnm\, (\mtil_k) = q_k - 1
\eqno(\new)
$$
where $q_k$ is defined in equation (\qj).
Equation (\ref1) is consistent with the power-law approximation used by the
code
since under this approximation the number distribution of particles within
the numerical mass bin $k$ is $n_k(m) = \rho_k(m)/m \propto m^{q_k - 1}$.

To determine the accuracy of the numerical results for $\dlnndlnm$,
we compare the numerical and analytic results for the three cases with exact
analytic solutions.
The analytic solutions are of the form $n \propto m^{-\tau} \exp[-b(t)\, m]$
or $\dlnndlnm = -\tau - b(t)\, m$ for $m \gg 1$, where $\tau = 0$, $3/2$,
and $5/2$ for $A_{ij} = 1$, $(i+j)/2$, and $ij$, respectively.
Thus $n \propto m^{-\tau}$ or $\dlnndlnm = -\tau$ for $1 \ll m \ll m_\ast$
when the characteristic mass $m_\ast(t)$ defined in equation (9) is large.
In figure 2(a) we show for each case the numerical and analytic $\dlnndlnm$ at
$m \lta m_\ast(t)$ at a given time.
As we noted in Paper I, there is a small lag in the evolution of the
numerical solutions for $A_{ij} = (i+j)/2$ and $ij$.
Therefore, in these cases, the numerical results are compared to the analytic
results at a slightly earlier time.
There are fluctuations in the numerical results in the first decade of the
logarithmically spaced mass bins ($1 \lta m/\Nbd \lta 10$) due to the
discreteness of the mass bins,
but the fluctuations are $\lta 0.015$.
The numerical results are much smoother and much more accurate outside this
mass range.
We can see from figure 2(a) that $\tau$ can be determined from the numerical
results at $1 \ll m \ll m_\ast$ to better than $\pm 0.001$.

Figure 2(b) is similar to figure 2(a), but it shows the mass range $m \gta
m_\ast(t)$.
The differences between the analytic results, which decrease linearly with
mass, and the numerical results are small,
but there is a small curvature in the numerical results, and the
numerical results become increasingly higher than the analytic results with
increasing mass.
(This is consistent with the observation in Paper I that the numerical
solutions show a slightly slower exponential decay at the high-mass end of
the mass spectrum.)
As a result, if we fit the numerical results near $\dlnndlnm = -20$ (or
$-30$) to a straight line $\dlnndlnm = -\theta - b m$, the resulting values
for $\theta$ are greater than the correct values (which are $\tau$ as given
above) by $0.13$--$0.15$ (or $0.29$--$0.36$).
This is the accuracy to which we can check whether a numerical solution is
consistent with $\dlnndlnm = -\theta - b m$ and a given $\theta$ at
$m \gg m_\ast$.

In Paper I we have discussed our numerical code in the context of numerical
codes in the astrophysics literature.
Other recent numerical codes include those by Krivitsky (1995), Hill and Ng
(1996), and Tzivion \etal\ (1999).
As we mentioned in Section 1, the numerical code used by Krivitsky does not
conserve mass and would have difficulty following the evolution of the mass
spectrum accurately for a long time.
The numerical code described by Hill and Ng conserves mass but uses a
relatively simple algorithm for distributing merger products.
We had in fact tried a similar algorithm for distributing merger products
(Quinlan and Shapiro 1989) before we developed the algorithm based on the
power-law approximation and had found that the high-mass end of the mass
spectrum converges very slowly with increasing grid resolution ($N_{bd}$)
if the rate coefficient increases steeply with the mass of the particles
(i.e., if $\nu$ and/or $\lambda$ is large).
Tzivion \etal\ have developed a mass-conserving numerical code that evolves
separately the total number ($N_k$) and mass ($M_k$) of particles in a
numerical mass bin $k$.
The numerical solutions obtained using this code appear to converge rapidly
with increasing grid resolution for the case $A_{m,m'} \propto m + m'$,
but there was no demonstration that this is also true for rate coefficients
with steeper mass dependence.

\sect{4.~RESULTS}

In this section we present the numerical solutions to the discrete form of
the coagulation equation for the rate coefficient and initial
conditions described in Section 2 (see also figure 1).
For the cases with $\nu \le 1$ (Sections 4.1--4.3), whenever possible, we
compare the properties of the numerical solutions to the predictions from
self-similar analysis.
Hereafter, unless otherwise stated, the self-similar analysis results cited
can be found in van Dongen and Ernst (1985a, 1988).

\subsec{4.1}{Orderly Growth Cases with $\nu \le 1$ and $\lambda < 1$}

An example of the numerical results for the mass spectrum evolution for
$\nu \le 1$ and $\lambda < 1$ is shown in figure 3.\footnote{$^3$}
{In figure 3 and all subsequent figures, the numerical mass spectrum plotted is
$n(\mtil_k) = N_k$ for $k \le \Nbd$ and $n(\mtil_k) = n_k(\mtil_k)$ for
$k > \Nbd$.}
In this and all other cases with $\nu \le 1$ and $\lambda < 1$, the
mass spectrum evolves in an orderly fashion.
For these cases, we stopped the numerical integrations when the asymptotic
behaviors of the solutions were clear and before $\Nbin$ reached $\Nmax$.

For orderly growth with $\nu \le 1$ and $\lambda < 1$, self-similar
analysis predicts that self-similar solutions have the form
$$
n_k(t) = m_\ast(t)^{-2} \varphi[m_k/m_\ast(t)]
\eqnam{\selfsimA}
\eqno(\new)
$$
where $\varphi(x)$ is a scaling function, and the characteristic mass
$m_\ast(t)$ scales as $t^{1/(1-\lambda)}$.
Different definitions of $m_\ast(t)$, which correspond to different scales
for $x = m_k/m_\ast(t)$ and $\varphi(x)$, can be used.
We adopt
$$
m_\ast(t) = {\cal M}_3(t)/{\cal M}_2(t)
\eqno(\new)
$$
where ${\cal M}_\ell(t) \equiv \sum_{k=1}^\infty m_k^\ell n_k(t)$ is the
$\ell$th moment of the mass spectrum.
This choice of $m_\ast$ is convenient because it can also be used for
runaway growth with $\nu \le 1$ and $\lambda > 1$ (Section 4.3).

In all cases, the numerical solution tends toward a self-similar solution
of the form (\selfsimA) at large $t$.
This is illustrated in figure 4, where we plot the numerical solution at
three different times in the form of $\log m^2 n$ as a function
of $\log m/m_\ast$ for the case shown in figure 3.
In the scaling limit (\selfsimA), $m^2 n = x^2 \varphi(x)$.
We have evaluated the exponent
$z(t_i) = \ln[m_\ast(t_{i+1})/m_\ast(t_{i-1})]/\ln(t_{i+1}/t_{i-1})$
from the numerical results for $m_\ast(t)$ at output times $t_{i-1}$,
$t_i$, and $t_{i+1}$ for all cases with $\nu \le 1$ and $\lambda < 1$.
In all but two cases, the exponent $z$ at large $t$ agrees with
$1/(1-\lambda)$ to better than one part in $1.5 \times 10^3$.
For the two cases with $(\mu,\nu) = (1/6,2/3)$ and $(1/3,1/2)$,
the agreement between the numerical results at the end of the numerical
runs, $z = 5.985$ and $5.963$, and the analytic result, $1/(1-\lambda) = 6$,
is slightly worse,
but the numerically determined exponents are still slowly increasing with
time at the end of the numerical runs, indicating that the numerical
results have not completely reached the asymptotic regime.

Figure 5 shows the numerical results for the scaling function $\varphi(x)$
for all cases with $\nu \le 1$ and $\lambda < 1$.
The location of the peak of $x^2 \varphi$ is not very sensitive to $\mu$ or
$\nu$, and it is at $x = 0.33$--$0.71$.
Self-similar analysis predicts that the scaling function $\varphi(x)$
decays exponentially at large $x$.
For $\nu < 1$, $\varphi(x) \propto x^{-\theta} \exp(-b x)$ or
$\dlnpdlnx = -\theta - b x$, with $\theta = \lambda$, at large $x$.
The detailed behavior of $\varphi(x)$ at large $x$ for
$\nu = 1$ depends on the specific form of $A_{ij}$.
For $A_{ij}$ in equation (\Aij) with $\nu =1$, the large-$x$ behavior of
$\varphi(x)$ is similar to that for $\nu < 1$, but $\theta = (\mu + 3)/2$
if $-1 \le \mu < 0$ (van Dongen 1987b; see also Ernst \etal\ 1984).
In all cases, the numerical $\varphi(x)$ decays exponentially at large $x$,
and $\dlnpdlnx$ at large $x$ is consistent with the analytic result (see
Section 3 for a discussion of the accuracy of the numerical results at
large $x$).

The behavior of the scaling function $\varphi(x)$ at small $x$ is
qualitatively different for $\mu < 0$, $\mu = 0$, and $\mu > 0$.
For the cases with $\mu < 0$ (figure 5(a)), $\varphi(x)$ also decays
exponentially at small $x$, because light particles are rapidly accreted
by heavy particles.
Self-similar analysis predicts that
$\varphi(x) \propto x^{-a} \exp(b x^\mu/\mu)$ or
$\dlnpdlnx = -a + b x^\mu$ at small $x$, where $a$ and $b$ are
constants that depend on the specific form of $A_{ij}$.
For $A_{ij}$ in equation (\Aij), $a = 2$ if $\nu > 0$ and $a = 1$ if $\nu = 0$.
In figure 6, we plot the numerical results for $\dlnpdlnx$ as a function of
$x^\mu$ for the cases with $\mu = -1/6$ to show that the numerical results
indeed have the form $\dlnpdlnx = -a + b x^\mu$ at large $x^\mu$ (or small
$x$).
Furthermore, in all cases with $\mu < 0$, the value of $a$ from
least-squares fit is in agreement with the analytic result to better
than $\pm 0.004$.

For the cases with $\mu = 0$, i.e., $A_{ij} = (i^\nu + j^\nu)/2$, the
scaling function shows a power-law behavior $\varphi(x) \propto x^{-\tau}$
at small $x$ (figure 5(b)).
The numerical results for the exponent $\tau$ are
$0.000$, $1.001$, $1.033$, $1.109$, $1.216$, $1.347$, and $1.500$ for
$\nu = 0$, $1/6$, $1/3$, $1/2$, $2/3$, $5/6$, and $1$, respectively.
Self-similar analysis gives $\tau = 2 - p_\lambda/w$, where $p_\lambda$ and
$w$ depend on the specific form of $A_{ij}$.
van Dongen and Ernst (1985b) have used this expression to derive analytic
lower and upper bounds on $\tau$ for $A_{ij} = (i^\nu + j^\nu)/2$.
Our numerical results are consistent with these bounds.
Note that the exponent $\tau$ is discontinuous at $\nu = 0$:
$\tau$ decreases from $1.5$ at $\nu = 1$ to $1$ as $\nu \downarrow 0$,
but $\tau = 0$ at $\nu = 0$
(recall that the $\nu = 0$ and $\nu = 1$ cases have exact analytic
solutions).
In contrast, for $A_{ij} \propto (i + j)^\nu$ (which also has $\mu = 0$),
the exponent $\tau$ decreases smoothly from $1.5$ at $\nu = 1$ to
$0$ at $\nu = 0$ (van Dongen and Ernst 1985c, Krivitsky 1995).

For the cases with $\mu > 0$ (figure 5(c)), the scaling function $\varphi(x)$
oscillates around a power law at small $x$, with the fractional amplitude
of the oscillation decreasing as $x \to 0$.
This damped oscillatory approach to a power law is shown more clearly in
figure 7, where we plot the logarithmic slope $\dlnpdlnx$ as a function of
$\log x$ for the cases with $\mu = 1/6$.
In all cases, the logarithmic slope tends to a constant value as
$x \to 0$, and the asymptotic value is consistent with the leading
small-$x$ behavior predicted by self-similar analysis:
$\varphi(x) \propto x^{-(1 + \lambda)}$ or $\dlnpdlnx = -(1 + \lambda)$.
The oscillation appears to be periodic in the variable $\ln x$, but it is
difficult to determine this accurately because of the limited number of
cycles seen in our numerical results.

The damped oscillatory behavior at small $x$ for $\mu > 0$ and
$\lambda < 1$ was not known previously.
In their self-similar analysis, van Dongen and Ernst (1988) were unable to
find higher-order corrections to the leading small-$x$ behavior of
$\varphi(x)$ for $\mu > 0$ and $\lambda < 1$, and they raised the
possibility that physically acceptable self-similar solutions may not exist.
As we have just shown, there are indeed physically acceptable self-similar
solutions, and they are reached from monodisperse initial conditions.
Based on our numerical results, we suggest that the leading small-$x$
behavior and the first correction could be of the form
$\varphi(x) \propto x^{-(1+\lambda)} [1 + f(x) \cos(B\ln x + C)]$,
where $f(x)$ is an increasing function of $x$, possibly $A x^\alpha$ with
$\alpha > 0$.
The failure to find the first correction in the self-similar analysis is
probably due to the unusual form of this correction.

As we mentioned in Section 1, Krivitsky (1995) has obtained numerical
solutions to the continuous form of the coagulation equation for
$A_{m,m'} \propto (m m')^{\lambda/2}$ with $\lambda \le 1$.
Krivitsky concluded that the numerical solutions are self-similar at large
time but that the asymptotic behavior at small mass is not a power law.
The latter conclusion is different from ours and, we believe, incorrect.
We can understand why Krivitsky reached this conclusion by examining
figure 5(b) of Krivitsky (1995), where the numerical results for $\dlnndlnm$
are shown for the case $\lambda = 0.4$.
By the last time shown, the numerical results have converged to a
self-similar form for $x \gta 10^{-4}$, and the logarithmic slope is indeed
decreasing with decreasing $x$ over the range $10^{-4} \lta x \lta 10^{-1}$.
However, as we can see from our figure 7, over this range in $x$, $\dlnpdlnx$
is in fact decreasing from a maximum to a minimum in its oscillatory
approach to a constant value.
Therefore, the incorrect conclusion was reached because Krivitsky did
not evolve the numerical solutions for a sufficiently long time to see the
true asymptotic behavior at small mass.

\subsec{4.2}{Orderly Growth Cases with $\nu \le 1$ and $\lambda = 1$}

On the borderline $\lambda = 1$ and $\nu \le 1$, the numerical mass
spectrum evolves in an orderly fashion, but with the characteristic mass
$m_\ast(t)$ increasing exponentially with time.
For these cases, we set $\Ndec = 19$ and stopped the numerical integrations
as soon as $\Nbin$ reached $\Nmax$, i.e., when the mass spectra extended
over $20$ decades in mass.

We distinguish the cases with $\mu = 0$ and $\mu > 0$.
As we discussed above, the case $\mu = 0$ is $A_{ij} = (i + j)/2$
with exact analytic solution, and the numerical solution for this case is
in excellent agreement with the analytic solution (see Section 3 and
figures 2 and 5(b)).
The mass spectrum tends toward a self-similar solution of the form
(\selfsimA) at large $t$, with $m_\ast(t) \propto e^t$ (see figure 8),
and the scaling function $\varphi(x) \propto x^{-3/2}$ at small $x$ and
$\propto x^{-\theta} \exp(-b x)$ with $\theta = (\mu + 3)/2 = 3/2$ at
large $x$.

For $\mu > 0$, van Dongen and Ernst (1988) have derived a modified
self-similar solution:
$$
n_k(t) = (m_\ast^2 \ln m_\ast)^{-1} \varphi(m_k/m_\ast)
\eqnam{\selfsimB}
\eqno(\new)
$$
where $(\ln m_\ast)^2 = a + b t$ and $a$ and $b$ are constants.
The scaling function $\varphi(x)$ is predicted to scale as $x^{-2}$ at
small $x$ and $x^{-1} \exp(-bx)$ at large $x$.

The numerical results for $m_\ast(t)$ for the three cases with $\mu > 0$
(and also the case $\mu = 0$) are shown in figure 8.
In each case, we have fitted the numerical results at the last two output
times to $\ln m_\ast = a + b t$ and $(\ln m_\ast)^2 = a + b t$, and they
are shown as dotted and solid lines, respectively.
For $(\mu,\nu) = (1/2,1/2)$, $(\ln m_\ast)^2 = a + b t$ provides a
reasonably good fit to the numerical results at large $t$.
For $(\mu,\nu) = (1/3,2/3)$ and, in particular, $(1/6,5/6)$, the numerical
results at large $t$ show deviations from $(\ln m_\ast)^2 = a + b t$.
The deviations in the last two cases are probably due to the numerical
results not having completely reached the asymptotic regime, but we cannot
rule out the possibility that the asymptotic behavior is different from the
analytic prediction.

If a numerical solution approaches the self-similar solution
(\selfsimB), we expect $m^2 n \ln m_\ast \to x^2 \varphi(x)$ and
$\dlnndlnm \to \dlnpdlnx$.
In figure 9 we show $\dlnndlnm$ as a function of $\log m/m_\ast$ at three
different times for all cases with $\mu > 0$ and $\lambda = 1$.
For $(\mu,\nu) = (1/2,1/2)$, $\dlnndlnm$ has converged to $\dlnpdlnx$
at $x = m/m_\ast \gta 10^{-5}$,
but the convergence at $x \lta 10^{-5}$ is very slow and is not complete by
the end of the numerical run.
The range of $x$ over which $\dlnndlnm$ has converged by the end of the
numerical run is wider for smaller $\mu$.
(A similar analysis of $m^2 n \ln m_\ast$ reveals a small increase in the
normalization of $m^2 n \ln m_\ast$ with time in the range of $x$ where
$\dlnndlnm$ has converged.
This increase is more pronounced for smaller $\mu$ and is probably due to
$m_\ast$ not having completely reached the asymptotic regime.)
For $(\mu,\nu) = (1/6,5/6)$, it is reasonably clear that $\dlnpdlnx \to
-2$ in the small-$x$ limit, consistent with the analytic prediction.
For $(\mu,\nu) = (1/3,2/3)$ and $(1/2,1/2)$, the small-$x$ behaviors are less
certain because of the slow convergence at small $x$, but they also appear
to be consistent with the analytic prediction.
Finally, in all three cases, the large-$x$ behavior of the numerical
$\dlnpdlnx$ is consistent with the analytic prediction that $\dlnpdlnx =
-1 - b x$.

\subsec{4.3}{Runaway Growth Cases with $\nu \le 1$ and $\lambda > 1$}

In all cases with $\nu \le 1$ and $\lambda > 1$, the numerical mass
spectrum develops in a finite time $t_0$ a power-law tail,
$n_k \propto k^{-\tau}$, at large mass that violates mass conservation, and
runaway growth is expected to start at $\tcrit = t_0$ in the limit
$n_0 \to \infty$.
For most cases with $\nu \le 1$ and $\lambda > 1$, we stopped the numerical
integrations as soon as $\Nbin$ reached $\Nmax$ and the mass spectra
extended over $20$ decades in mass;
so the numerical solutions approach very close to but do not exceed $t_0$.
To study the transition at $t = t_0$, we extended the integrations for the
cases $(\mu,\nu) = (1/3,1)$ and $(2/3,2/3)$ to $t > t_0$.
Figure 10 shows the numerical results for the mass spectrum evolution for
the case $(\mu,\nu) = (1/3,1)$ at $t \le t_0$ (solid lines) and $t > t_0$
(dashed lines).

For runaway growth with $\nu \le 1$ and $\lambda > 1$, self-similar
analysis predicts that self-similar solutions close to but before $t_0$
have the form
$$
n_k(t) = m_\ast(t)^{-\tau} \varphi[m_k/m_\ast(t)]
\eqnam{\selfsimC}
\eqno(\new)
$$
where the scaling function $\varphi(x) \propto x^{-\tau}$ at small $x$,
the characteristic mass $m_\ast(t)$ diverges as $(t_0 - t)^{-1/\sigma}$,
and $\sigma = \lambda + 1 - \tau$.
With $m_\ast(t)$ diverging at $t_0$, the self-similar solution
(\selfsimC) has $n_k(t_0) \propto k^{-\tau}$ at large $k$.
In all cases, the numerical solution tends toward a self-similar solution
of the form (\selfsimC) as $t \uparrow t_0$, and the numerical $\varphi(x)$
is indeed a power law at small $x$.
This is illustrated in figure 11, where we plot the numerical solution at
three different times ($< t_0$) in the form of
$\log(m^2 n\, m_\ast^{\tau-2})$ as a function of $\log m/m_\ast$ for the
case shown in figure 10, with the exponent $\tau$ determined from the
numerical solution itself (see table 1 and discussion below).
In the scaling limit (\selfsimC), $m^2 n\, m_\ast^{\tau-2} = x^2
\varphi(x)$.

Despite the change in the form of the self-similar solution, the analytic
predictions for the large-$x$ behavior of $\varphi(x)$ are similar to those
for orderly growth in Sections 4.1--4.2:
$\varphi(x) \propto x^{-\theta} \exp(-b x)$ or $\dlnpdlnx = -\theta - b x$,
where $\theta = \lambda$ if $\nu < 1$, $\theta = (\mu + 3)/2$ if $\nu = 1$
and $0 < \mu < 1$, and $\theta = 5/2$ if $\nu = \mu = 1$.
In all cases, the numerical $\varphi(x)$ decays exponentially at large $x$,
and $\dlnpdlnx$ at large $x$ is consistent with the analytic result.

The analytic predictions for the exponents $\tau$ and $\sigma$ are $\tau =
(\lambda + 3)/2$ and $\sigma = \lambda + 1 - \tau = (\lambda - 1)/2$
(Hendriks \etal\ 1983, van Dongen and Ernst 1985a, 1988).
For comparison, we have determined $\tau$ and $\sigma$ from the numerical
results for $\dlnpdlnx$ and $m_\ast(t)$, respectively.
In figure 12 we show the numerical results for $\dlnpdlnx$ for the cases
with $\lambda = 4/3$.
The numerical results clearly converge to constant values at small $x$.
The constant asymptotic values are consistent with $\varphi(x) \propto
x^{-\tau}$ or $\dlnpdlnx = -\tau$ at small $x$ and directly yield the
values of $\tau$.
It is also clear from figure 12 that the asymptotic values and hence $\tau$
for these cases with the same $\lambda$($= 4/3$) are different from each
other and from the analytic prediction that
$\dlnpdlnx = -\tau = -(\lambda + 3)/2 = -13/6$ (dashed line in figure 12).
The numerical results for the exponent $\tau$ for all cases with
$\nu \le 1$ and $\lambda > 1$ are listed in table 1.
The only case where the exponent $\tau$ agrees with the analytic prediction
$(\lambda + 3)/2$ is the case $\nu = \mu = 1$ (i.e., the case $A_{ij} = ij$
with exact analytic solution).
In all other cases, the exponent $\tau$ is less than $(\lambda + 3)/2$.
For a given $\lambda$, the deviation of $\tau$ from $(\lambda + 3)/2$ is
largest for $\nu = 1$ and smallest for $\nu = \mu$.

To determine the exponent $\sigma$, we fit the numerical results for
$m_\ast(t)$ at output times $t_{i-1}$, $t_i$, and $t_{i+1}$ to $m_\ast(t) =
C (t_0' - t)^{-1/\sigma'}$ to obtain $C(t_i)$, $t_0'(t_i)$, and
$\sigma'(t_i)$.
In most cases, $\sigma'(t)$ has converged to a constant value by the end of
the numerical run and directly yields $\sigma$.
In the remaining cases, we extrapolated $\sigma'(t)$ to the limit
$m_\ast(t) \to \infty$ to obtain $\sigma$, but the difference between
$\sigma'(t)$ at the end of the numerical run and $\sigma$ is $\le 0.002$.
The numerical results for the exponent $\sigma$ are listed in table 1,
together with $\lambda + 1 - \tau$.
Except for the case $\nu = \mu = 1$, the exponent $\sigma$ is greater than
the analytic prediction $(\lambda - 1)/2$.
For a given $\lambda$, the deviation of $\sigma$ from $(\lambda - 1)/2$ is
largest for $\nu = 1$ and smallest for $\nu = \mu$.
We note that the numerical results for $\sigma$ and $\tau$ are
consistent with one another in that they satisfy the relation $\sigma =
\lambda + 1 - \tau$ for self-similar solutions of the form (\selfsimC)
to within $\pm 0.001$.

The procedure described above for determining the exponent $\sigma$ also
yields the time $t_0$.
The numerical results for $t_0$, in units of $1/(n_0 A_{11})$, are listed
in table 1.
Since we expect $t_0 \sim 1/[(\lambda-1) n_0 A_{11}]$ (see Paper I), it is
convenient to parameterize $t_0$ as $t_0 = K/[(\lambda-1) n_0 A_{11}]$.
The parameter $K$ is shown in figure 13.
We find that $K = 1$--$2$ if $\lambda \gta 1.1$ and that, for a given
$\lambda$, $K$ is smallest for $\nu = 1$ and largest for $\nu = \mu$.
For $\nu = 1$, the parameter $K$ shows a maximum at $\lambda \approx 1.3$.
For $\nu = \mu$, $K$ increases monotonically with decreasing $\lambda$,
but it is unclear whether $K$ approaches a constant value or diverges
as $\lambda \to 1$.
Finally, we note that the numerical results for $t_0$ are consistent with
the bound $t_0 \ge 1/[(\lambda - 1) n_0 A_{11}]$ for $A_{ij}$ in
equation (\Aij)
and monodisperse initial conditions and the stronger bound $t_0 \ge
1/[(2^{\lambda - 1} - 1) n_0 A_{11}]$ for $\nu = \mu$, derived analytically
by Hendriks \etal\ (1983).

We have found that the exponents $\tau$ and $\sigma$ are in general
different from the analytic predictions.
Since the analytic prediction for $\sigma$ follows from that for $\tau$
and the relation $\sigma = \lambda + 1 - \tau$ for self-similar solutions
of the form (\selfsimC), and the numerical results for $\sigma$ and $\tau$
satisfy this relation, we have essentially a discrepancy in the exponent
$\tau$.
Let us examine the derivation of the analytic prediction for $\tau$, which
can be summarized as follows (see Hendriks \etal\ 1983, van Dongen and
Ernst 1985a, 1988 for details).
The mass flux from particles of mass $m_i \le m_k$ to particles of
mass $m_i > m_k$ is
$$
F_k(t) = \sum_{i=1}^k \sum_{j=k+1-i}^\infty m_i A_{ij} n_i(t) n_j(t) .
\eqno(\new)
$$
It is assumed that solutions to the coagulation equation at $t > t_0$
violate mass conservation by having a non-zero and finite mass flux to the
infinite-mass bin (i.e., in the limit $k \to \infty$), which is possible
only if the mass spectrum at $t > t_0$ is of the form $n_k(t) \propto
k^{-\tau'}$ at large $k$.
With $n_k(t) \propto k^{-\tau'}$ at large $k$, the mass flux
$F_k(t) \propto k^{\lambda + 3 - 2\tau'}$ at large $k$.
Thus $\tau' = (\lambda + 3)/2$ if the mass flux $F_k(t)$ is required to
be non-zero and finite in the limit $k \to \infty$.
Since the self-similar solution (\selfsimC) has $n_k(t_0) \propto
k^{-\tau}$ at large $k$, $\tau = \tau' = (\lambda + 3)/2$ if we assume that
the large-$k$ behavior at $t > t_0$ is also valid at $t = t_0$.
It should be noted that the arguments leading to $\tau = (\lambda + 3)/2$
are not rigorous.
In particular, as van Dongen and Ernst (1988) pointed out, one cannot
exclude the possibility that the mass flux diverges at one instant of time,
i.e., $t_0$.
We have found that $\tau < (\lambda + 3)/2$ in general.
This implies that the mass flux $F_k$ at $t_0$, which is $\propto
k^{\lambda + 3 - 2\tau}$ at large $k$, diverges as $k \to \infty$.
Thus the numerical solutions violate mass conservation at $t_0$ by having a
diverging mass flux to the infinite-mass bin.

Since the mass flux cannot diverge for all times $t > t_0$,
do the solutions at $t > t_0$ have the form $n_k(t) \propto
k^{-(\lambda + 3)/2}$ at large $k$ for non-zero and finite mass flux to
the infinite-mass bin?
If so, how can this large-$k$ behavior at $t > t_0$ be reconciled with
that at $t = t_0$?
To answer these questions, we have extended the numerical integrations
for the cases $(\mu,\nu) = (1/3,1)$ and $(2/3,2/3)$ to $t > t_0$.
The results for the mass spectrum evolution at $t > t_0$ for the case
$(\mu,\nu) = (1/3,1)$ are shown as dashed lines in figure 10.
As in the test case $A_{ij} = ij$ discussed in Section 3, the numerical
solution at $m_k \sim \mmax$ is affected by the finite value of the
maximum particle mass, $\mmax$, in the computational grid.
Otherwise, the mass spectrum at $t > t_0$ is indeed of the form
$n_k(t) \propto k^{-(\lambda + 3)/2}$ at large $k$, with the value of the
exponent confirmed by an analysis of the logarithmic slope.
Note, however, that the range of $k$ where $n_k(t) \propto
k^{-(\lambda + 3)/2}$ shrinks toward infinity as $t \downarrow t_0$.
The numerical solution for the case $(\mu,\nu) = (2/3,2/3)$ shows the same
large-$k$ behaviors at $t > t_0$.
Therefore, we conclude that the transition at $t = t_0$ is accomplished as
follows.
As $t \uparrow t_0$, the solution tends toward a self-similar solution of
the form (\selfsimC), with $n_k(t) \propto k^{-\tau}$ for $1 \ll m_k \ll
m_\ast(t)$, $\tau < (\lambda + 3)/2$ in general, and $m_\ast(t) \to \infty$.
As $t \downarrow t_0$, $n_k(t) \propto k^{-(\lambda + 3)/2}$ for
$m_k \gg m'_\ast(t)$ and $m'_\ast(t) \to \infty$.

\subsec{4.4}{Runaway Growth Cases with $\nu > 1$}

Examples of the numerical results for the mass spectrum evolution with
$\nmin = 10^{-30}$ (solid lines) and $10^{-35}$ (dotted lines) for
$\nu > 1$ are shown in the lower panels of figure 14.
In all cases with $\nu = 2$ and $3/2$ and the three cases with $\nu = 7/6$
and $\mu \le 0$ (see, e.g., figure 14(a)), the behaviors of the numerical
solutions for both $\nmin = 10^{-30}$ and $10^{-35}$ are qualitatively
similar to those found in Paper I for other forms of $A_{ij}$ with $\nu > 1$.
In the early stages, the mass spectrum appears to decay exponentially at
large mass.
After some evolution, the mass spectrum shows an exponential drop in an
intermediate mass range and a power-law tail of the form $n_k \propto
k^{-\nu}$ (or $n \propto m^{-\nu}$) at large mass.
For the three cases with $\nu = 7/6$ and $\mu \ge 1/3$, the numerical
solutions with $\nmin = 10^{-30}$ do not develop the $m^{-\nu}$ tail when
$\Nbin$ reaches $\Nmax$ (see, e.g., figure 14(b)), and their behaviors are
qualitatively similar to those for the runaway growth cases with $\nu \le 1$
and $\lambda > 1$ (Section 4.3; figure 10).
When $\nmin$ is decreased to $10^{-35}$, the $\mu = 1/3$ case does develop
an $m^{-\nu}$ tail when $\Nbin$ reaches $\Nmax$ (figure 14(b)), but the other
two cases do not.
It is not feasible to compute solutions for much smaller $\nmin$,
but we strongly suspect that the remaining two cases would develop the
$m^{-\nu}$ tail for sufficiently small $\nmin$.

The fact that the tail at large mass is of the form $n \propto m^{-\nu}$
or $\dlnndlnm = -\nu$ is illustrated in the upper panels of figure 14,
where we plot the numerical results for $\dlnndlnm$ at the specified times,
just after the formation of the tail for the runs with $\nmin = 10^{-35}$.
As found in Paper I, the time at which the power-law tail develops
decreases slowly toward zero as $\nmin$ decreases (lower panels of figure 14).
We do not repeat the demonstrations here, but it was shown in Paper I that
numerical solutions with different maximum particle mass, $\mmax$, included
in the computational grid are identical in the overlapping mass range and
that the power-law tail simply extends to larger particle mass $m$ when
$\mmax$ is increased.
(It was also shown that the numerical solutions converge by $\Nbd = 40$ and
$\delta_M = 5\%$.)
Therefore, in the limit $\nmin \to 0$ and $\mmax \to \infty$, the numerical
solutions for all cases with $\nu > 1$ (with the possible exception of the
cases with $\nu = 7/6$ and $\mu > 1/3$) should develop in an infinitesimal
time power-law tails of the form $n \propto m^{-\nu}$ that extend to
arbitrarily large $\mmax$.
However, this mass spectrum is not self-consistent because the power-law
tail implies a mass flux and, if $1 < \nu \le 2$, a cumulative mass that
diverge with $\mmax$ (see footnote 2 and Paper I).
Thus, as in Paper I, the numerical results strongly suggest that there are
no self-consistent solutions to the coagulation equation at any time if
$\nu > 1$.
Furthermore, since the formation of the power-law tail in the coagulation
equation solution with finite $\nmin$ probably corresponds to the onset of
runaway growth in Monte Carlo simulations with finite $n_0$ (see Paper I)
and decreasing $\nmin$ is equivalent to increasing $n_0$ (see Section 3),
the time $\tcrit$ for the onset of runaway growth for $\nu > 1$ should
decrease slowly toward zero with increasing $n_0$.

\sect{5.~CONCLUSIONS}

We have conducted a systematic survey of numerical solutions to the
coagulation equation (\dCoEq) for a rate coefficient of the form
$A_{ij} \propto (i^\mu j^\nu + i^\nu j^\mu)$ and monodisperse initial
conditions.
The numerical results confirmed that there are three classes of rate
coefficients with qualitatively different solutions to the coagulation
equation.

For $\nu \le 1$ and $\lambda = \mu + \nu < 1$ (Section 4.1), the numerical
solution evolves in an orderly fashion and tends toward a self-similar
solution of the form (\selfsimA) at large time $t$.
The time dependence of the characteristic mass $m_\ast(t)$ at large $t$ and
the behaviors of the scaling function $\varphi(x)$ at small and large
$x = m_k/m_\ast(t)$ agree with the analytic predictions of van Dongen and
Ernst (1985a, 1988).
In particular, for the subset of cases with $\mu > 0$, we disagreed with
the earlier numerical study by Krivitsky (1995) and found that $\varphi(x)$
does in fact approach the analytically predicted power-law behavior
$\varphi(x) \propto x^{-(1+\lambda)}$ at small $x$, but in a damped
oscillatory fashion that was not known previously (figure 7).
For $\mu = 0$, we determined the exponent $\tau$ of the power-law behavior
$\varphi(x) \propto x^{-\tau}$ at small $x$.

On the borderline $\nu \le 1$ and $\lambda = 1$ (Section 4.2), the numerical
solution evolves in an orderly fashion.
For $\mu = 0$, i.e., $A_{ij} \propto i + j$, the numerical solution is in
excellent agreement with the exact analytic solution and tends toward a
self-similar solution of the form (\selfsimA) at large $t$.
For $\mu > 0$, the numerical solution appears to tend toward a self-similar
solution of the form (\selfsimB), but we had limited success in
comparing the behaviors of $m_\ast(t)$ and $\varphi(x)$ at small $x$ to the
analytic predictions because the convergence to the self-similar solution
is very slow.

For $\nu \le 1$ and $\lambda > 1$ (Section 4.3), the numerical mass
spectrum $n_k$
develops in a finite time $t_0$ a power-law tail, $n_k \propto k^{-\tau}$,
at large $k$ that violates mass conservation, and runaway growth/gelation
is expected to start at $\tcrit = t_0$ in the limit that the initial
number of particles $n_0 \to \infty$.
As $t \uparrow t_0$, the numerical solution tends toward a self-similar
solution of the form (\selfsimC), with $\varphi(x) \propto x^{-\tau}$ at small
$x$ and $m_\ast(t)$ diverging as $(t_0 - t)^{-1/\sigma}$.
The exponent $\tau$ is in general less than the analytic prediction
$(\lambda + 3)/2$, and the exponent $\sigma$ greater
than the analytic prediction $(\lambda - 1)/2$, but they satisfy the
relation $\sigma = \lambda + 1 - \tau$ for self-similar solutions of the
form (\selfsimC) (table 1).
We studied the dependence of $t_0$ on the exponents $\mu$, $\nu$, and
$\lambda$ and found that $t_0 = K/[(\lambda-1) n_0 A_{11}]$ and
$K = 1$--$2$ if $\lambda \gta 1.1$ (table 1; figure 13).
At $t > t_0$, $n_k \propto k^{-(\lambda + 3)/2}$ for
$m_k \gg m'_\ast(t)$, with $m'_\ast(t) \to \infty$ as $t \downarrow t_0$.

For $\nu > 1$ (Section 4.4), the behaviors of the numerical solution are
qualitatively similar to those found in Paper I:
the numerical mass spectrum develops a power-law tail of the form $n_k \propto
k^{-\nu}$ at large $k$ that is not self-consistent, and the time at which
the power-law tail develops decreases toward zero as the numerical
parameter $\nmin$ decreases.
The numerical results strongly suggest that there are no self-consistent
solutions to the coagulation equation at any time and that runaway
growth/gelation is instantaneous in the limit $n_0 \to \infty$.
They also indicate that the time $\tcrit$, in units of $1/(n_0 A_{11})$,
for the onset of runaway growth decreases slowly toward zero with
increasing $n_0$, consistent with recent Monte Carlo simulation results
(Malyshkin and Goodman 2001).

The results presented in this paper and Paper I suggest several problems
for future investigations.
First, as we pointed out in Section 4.1, the failure to find the first
correction to the leading small-$x$ behavior $\varphi(x) \propto
x^{-(1+\lambda)}$ for the orderly growth cases with $\mu > 0$ and
$\lambda < 1$ in previous self-similar analysis is probably due to the
unusual, damped oscillatory form of this correction.
Given the information provided by the numerical results, it may now be
possible to derive the first correction analytically.
We suggested that the first correction could, e.g., be of the form
$x^{-(1+\lambda)} f(x) \cos(B \ln x + C)$, where $f(x)$ is an increasing
function of $x$.
Second, we have found that the exponent $\tau$ for the
runaway growth cases with $\nu \le 1$ and $\lambda > 1$ is in general
different from existing analytic prediction.
It is important to investigate whether analytic prediction matching the
numerical results could be derived.
In this connection, it would be useful to obtain numerical solutions for
other forms of $A_{ij}$ and determine whether the exponent depends on the
specific form of $A_{ij}$.
Finally, it should be emphasized that rate coefficients with $\nu > 1$ do
arise, and are of great interest, in astrophysics (see, e.g., Lee 1993,
2000, and references therein).
For astrophysical applications, we are interested in systems with finite
$n_0$ and interactions with the runaway particle.
Thus, for $\nu > 1$, a detailed comparison of numerical solutions with finite
$\nmin$ and Monte Carlo simulations with finite $n_0$ should be conducted
to test the correspondence between them, and
the question of whether the coagulation equation can be modified to take
into account the interactions between the runaway particle and the other
particles should also be investigated.

\vfill\eject

\sect{ACKNOWLEDGEMENTS}

I thank Martin Duncan, Stan Peale, and the referees for their helpful
comments on the manuscript.
This work was supported in part by the NASA PG\&G Program under
Grant NAG5-3646.

\vfill\eject

\def\dline{\noalign{\hrule \vskip 2pt \hrule \vskip 10pt}}
\def\cline{\noalign{\vskip 4pt\hrule \vbox to 3pt{}}}
\def\bline{\noalign{\vskip 4pt\hrule}}

\baselineskip=20pt

\centerline{{\bf Table 1.}\ \ Runaway growth cases with $\nu \le 1$ and
            $\lambda > 1$.}
\vskip -12pt
$$\vbox{
\halign{$\quad\hfil#$&$\qquad\hfil#$&$\qquad\hfil#$&$\qquad\hfil#\hfil$&$\qquad\hfil#\hfil$&$\qquad\hfil#\hfil$&$\qquad\hfil#\quad$\cr
\dline
\lambda~&\mu~&\nu~&\tau&\sigma&\lambda+1-\tau&n_0 A_{11} t_0\cr
\cline
 7/6&  1/6&    1&2.012&0.154&0.155& 7.136~\cr
 7/6&  1/3&  5/6&2.038&0.128&0.129& 9.146~\cr
 7/6&  1/2&  2/3&2.054&0.112&0.113&10.633~\cr
 7/6& 7/12& 7/12&2.057&0.110&0.110&10.855~\cr
 4/3&  1/3&    1&2.076&0.257&0.257& 3.741~\cr
 4/3&  1/2&  5/6&2.103&0.230&0.230& 4.284~\cr
 4/3&  2/3&  2/3&2.112&0.221&0.221& 4.492~\cr
 3/2&  1/2&    1&2.166&0.334&0.334& 2.451~\cr
 3/2&  2/3&  5/6&2.184&0.316&0.316& 2.639~\cr
 3/2&  3/4&  3/4&2.186&0.314&0.314& 2.664~\cr
 5/3&  2/3&    1&2.269&0.398&0.398& 1.750~\cr
 5/3&  5/6&  5/6&2.276&0.390&0.391& 1.808~\cr
11/6&  5/6&    1&2.380&0.453&0.453& 1.307~\cr
11/6&11/12&11/12&2.381&0.451&0.452& 1.317~\cr
   2&    1&    1&2.500&0.500&0.500& 1.000~\cr
\bline
}}$$

\baselineskip=13pt

\vfill\eject

\sect{REFERENCES}

\refpaper{Bak T A and Heilmann O J 1994}
         {Post-gelation solutions to Smoluchowski's coagulation equation}
         {J.\ Phys.\ A: Math.\ Gen.}{27}{4203--4209}
\refbook{Drake R L 1972}
        {A general mathematical survey of the coagulation equation
         {\it Topics in Current Aerosol Research (Part 2)} ed G M Hidy
         and J R Brock}
        {(Oxford: Pergamon) pp 201--376}
\refbook{Ernst M H 1986}
        {Kinetics of clustering in irreversible aggregation
         {\it Fractals in Physics} ed L Pietronero and E Tosatti}
        {(Amsterdam: North-Holland) pp 289--302}
\refpaper{Ernst M H, Hendriks E M and Leyvraz F 1984}
         {Smoluchowski's equation and the $\theta$-exponent for branched
          polymers}{J.\ Phys.\ A: Math.\ Gen.}{17}{2137--2144}
\refpaper{Hendriks E M, Ernst M H and Ziff R M 1983}
         {Coagulation equations with gelation}{J.\ Stat.\ Phys.}{31}{519--563}
\refpaper{Hill P J and Ng K M 1996}
         {New discretization procedure for the agglomeration equation}
         {AIChE J.}{42}{727--741}
\refpaper{Jeon I 1998}
         {Existence of gelling solutions for coagulation-fragmentation
          equations}{Commun.\ Math.\ Phys.}{194}{541--567}
\refpaper{Jeon I 1999}
         {Spouge's conjecture on complete and instantaneous gelation}
         {J.\ Stat.\ Phys.}{96}{1049--1070}
\refbook{Jullien R and Botet R 1987}
        {{\it Aggregation and Fractal Aggregates}}
        {(Singapore: World Scientific)}
\refpaper{Krivitsky D S 1995}
         {Numerical solution of the Smoluchowski kinetic equation and
          asymptotics of the distribution function}
         {J.\ Phys.\ A: Math.\ Gen.}{28}{2025--2039}
\apj{Lee M H 1993}{$N$-body evolution of dense clusters of compact stars}
    {418}{147--162}
\icarus{Lee M H 2000}
       {On the validity of the coagulation equation and
        the nature of runaway growth}{143}{74--86}
\refpaper{Leyvraz F and Tschudi H R 1981}
         {Singularities in the kinetics of coagulation processes}
         {J.\ Phys.\ A: Math.\ Gen.}{14}{3389--3405}
\icarus{Malyshkin L and Goodman J 2001}
       {The timescale of runaway stochastic coagulation}
       {150}{314--322}
\refpaper{McLeod J B 1962}
         {On a recurrence formula in differential equations}
         {Quart.\ J.\ Math.\ Oxford}{13}{283--284}
\apj{Quinlan G D and Shapiro S L 1989}
    {Dynamical evolution of dense clusters of compact stars}{343}{725--749}
\refpaper{Spouge J L 1985}
         {Monte Carlo results for random coagulation}
         {J.\ Colloid Interface Sci.}{107}{38--43}
\refpaper{Trubnikov B A 1971}
         {Solution of the coagulation equations in the case of a bilinear
          coefficient of adhesion of particles}{Sov.\ Phys.\ Dokl.}{16}
         {124--126}
\refpaper{Tzivion S, Reisin T G and Levin Z 1999}
         {A numerical solution of the kinetic collection equation using
          high spectral grid resolution: A proposed reference}
         {J.\ Comput.\ Phys.}{148}{527--544}
\refpaper{van Dongen P G J 1987{\rm a}}
         {On the possible occurrence of instantaneous gelation in
          Smoluchowski's coagulation equation}
         {J.\ Phys.\ A: Math.\ Gen.}{20}{1889--1904}
\refpaper{van Dongen P G J 1987{\rm b}}
         {Solutions of Smoluchowski's coagulation equation at large
          cluster sizes}{Physica A}{145}{15--66}
\refpaper{van Dongen P G J and Ernst M H 1985{\rm a}}
         {Dynamic scaling in the kinetics of clustering}
         {Phys.\ Rev.\ Lett.}{54}{1396--1399}
\refpaper{van Dongen P G J and Ernst M H 1985{\rm b}}
         {Comment on ``Large-time behavior of the Smoluchowski equations
          of coagulation''}{Phys.\ Rev.\ A}{32}{670--672}
\refpaper{van Dongen P G J and Ernst M H 1985{\rm c}}
         {Cluster size distribution in irreversible aggregation at large time}
         {J.\ Phys.\ A: Math.\ Gen.}{18}{2779}
\refpaper{van Dongen P G J and Ernst M H 1988}
         {Scaling solutions of Smoluchowski's coagulation equation}
         {J.\ Stat.\ Phys.}{50}{295--329}
\icarus{Wetherill G W 1990}
       {Comparison of analytical and physical modeling of planetesimal
        accumulation}{88}{336--354}
\refpaper{Ziff R M 1980}
         {Kinetics of polymerization}{J.\ Stat.\ Phys.}{23}{241--263}
\refpaper{Ziff R M, Ernst M H and Henriks E M 1983}
         {Kinetics of gelation and universality}
         {J.\ Phys.\ A: Math.\ Gen.}{16}{2293--2320}

\vfill\eject

\input epsf.tex

\epsfxsize=5.5truein \epsfbox[18 144 576 702]{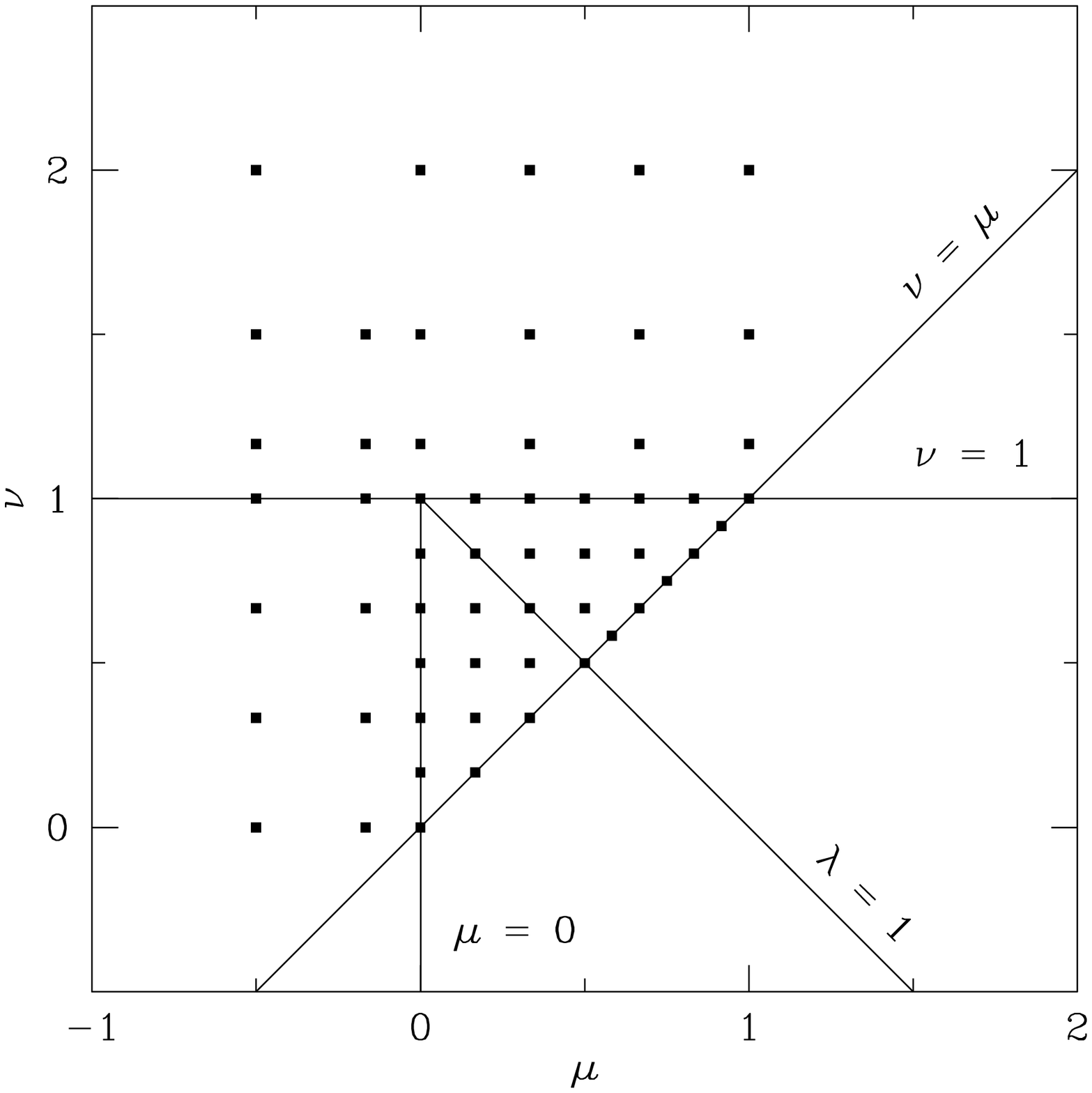}

{\bf Figure 1.}\ \ The exponents $\mu$ and $\nu$ of the cases for which we
have obtained numerical solutions to the discrete form of the coagulation
equation with rate coefficient $A_{ij} = (i^\mu j^\nu + i^\nu j^\mu)/2$.
The orderly growth cases with $\nu \le 1$ and $\lambda < 1$ and on the
borderline $\nu \le 1$ and $\lambda = 1$ are discussed in Sections 4.1 and
4.2, respectively.
The runaway growth cases with $\nu \le 1$ and $\lambda > 1$ are discussed
in Section 4.3, and those with $\nu > 1$ are discussed in Section 4.4.

\vfill\eject

\epsfxsize=5truein \epsfbox[72 36 594 684]{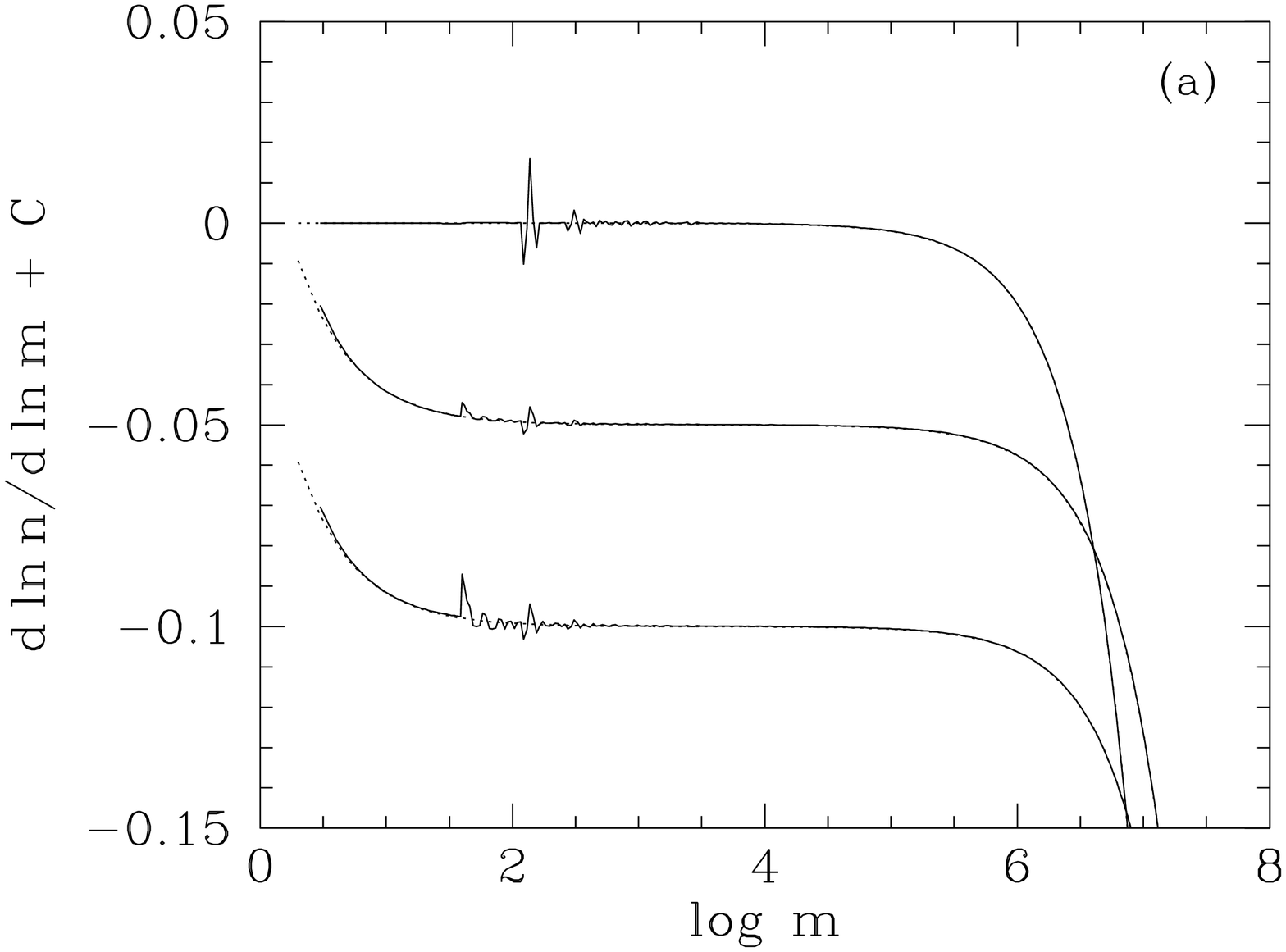}

\vfill\eject

\epsfxsize=5truein \epsfbox[72 36 594 684]{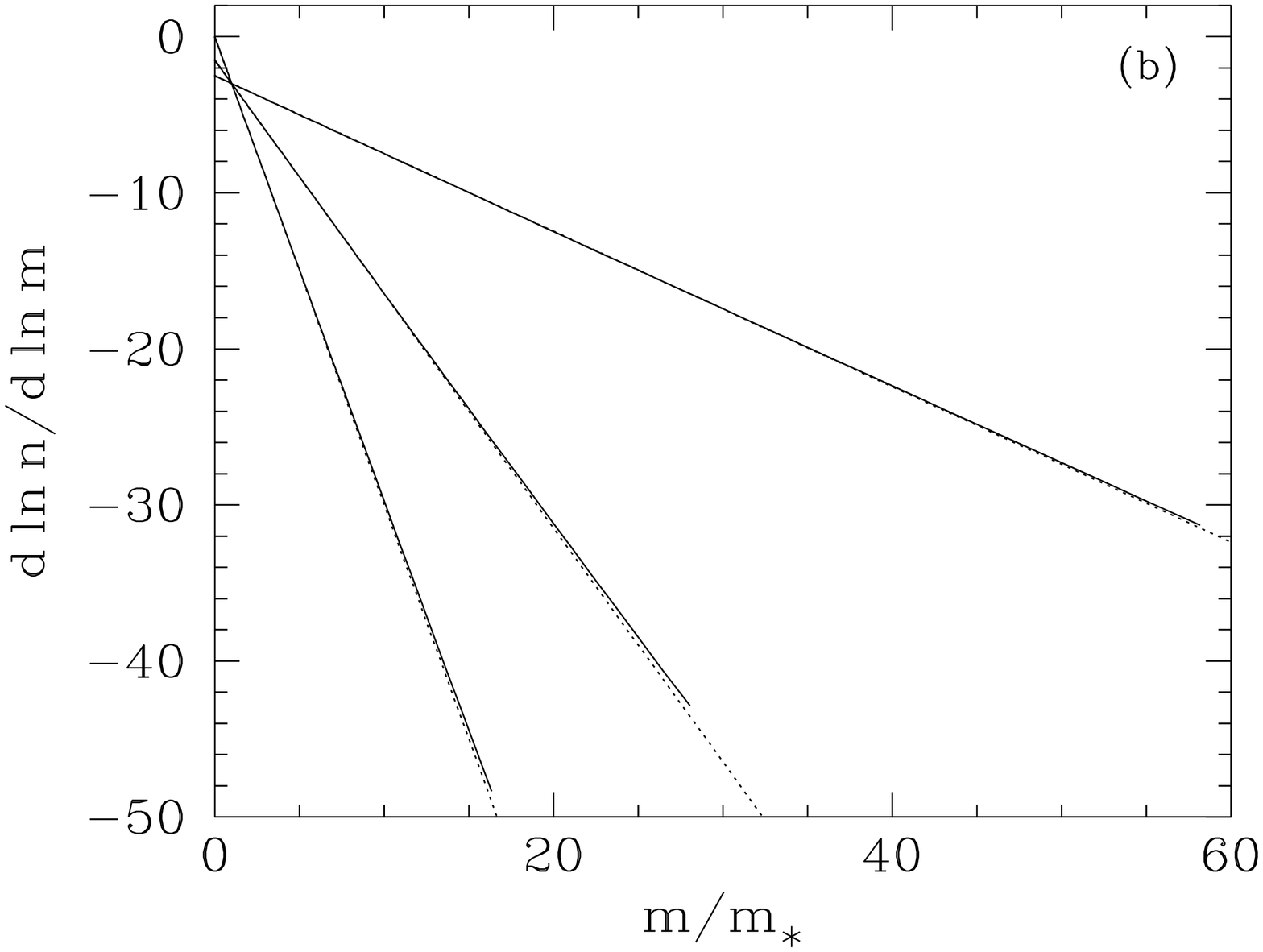}

{\bf Figure 2.}\ \ (a) Comparison of numerical (solid lines) and analytic
(dotted lines) results for $\dlnndlnm$ at $m \lta m_\ast(t)$.
The lines, from top to bottom, are $\dlnndlnm + C$ for $A_{ij} = 1$ at
$t = 10^8$, $A_{ij} = (i+j)/2$ at $t = 18$, and $A_{ij} = ij$ at $t = 1$;
the constant $C = 0$, $1.45$, and $2.4$, respectively.
For $A_{ij} = (i+j)/2$ and $ij$, there is a small lag in the evolution of
the numerical solutions, and the numerical results are compared to the
analytic results at a slightly earlier time $(1-\epsilon) t$, where
$\epsilon = 3.5 \times 10^{-4}$ and $1.1 \times 10^{-4}$, respectively.
(b) Same as (a), but at $m \gta m_\ast(t)$.
The lines, in decreasing steepness, are $\dlnndlnm$ for $A_{ij} = 1$ at
$t = 10^8$, $A_{ij} = (i+j)/2$ at $t = 18$, and $A_{ij} = ij$ at $t = 1$.

\vfill\eject

\epsfxsize=5truein \epsfbox[72 36 594 684]{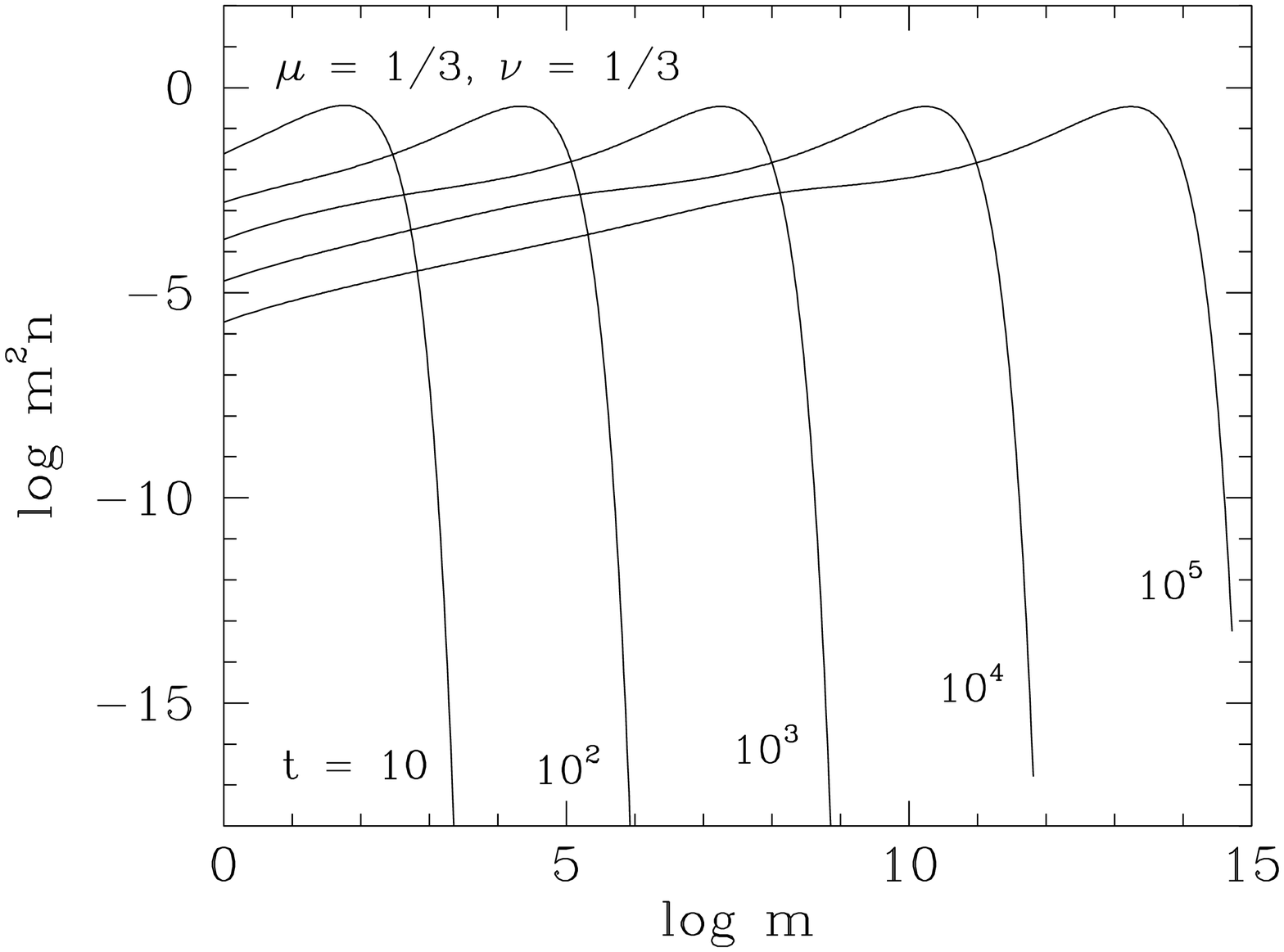}

{\bf Figure 3.}\ \ Evolution of the mass spectrum for $(\mu,\nu) = 
(1/3,1/3)$.
We plot $\log m^2 n$ as a function of $\log m$ since $m^2 n$ is the total
mass per unit logarithmic mass interval:
$\int m^2 n\,d\ln{m} = \int m n\,dm$.

\vfill\eject

\epsfxsize=5truein \epsfbox[72 36 594 684]{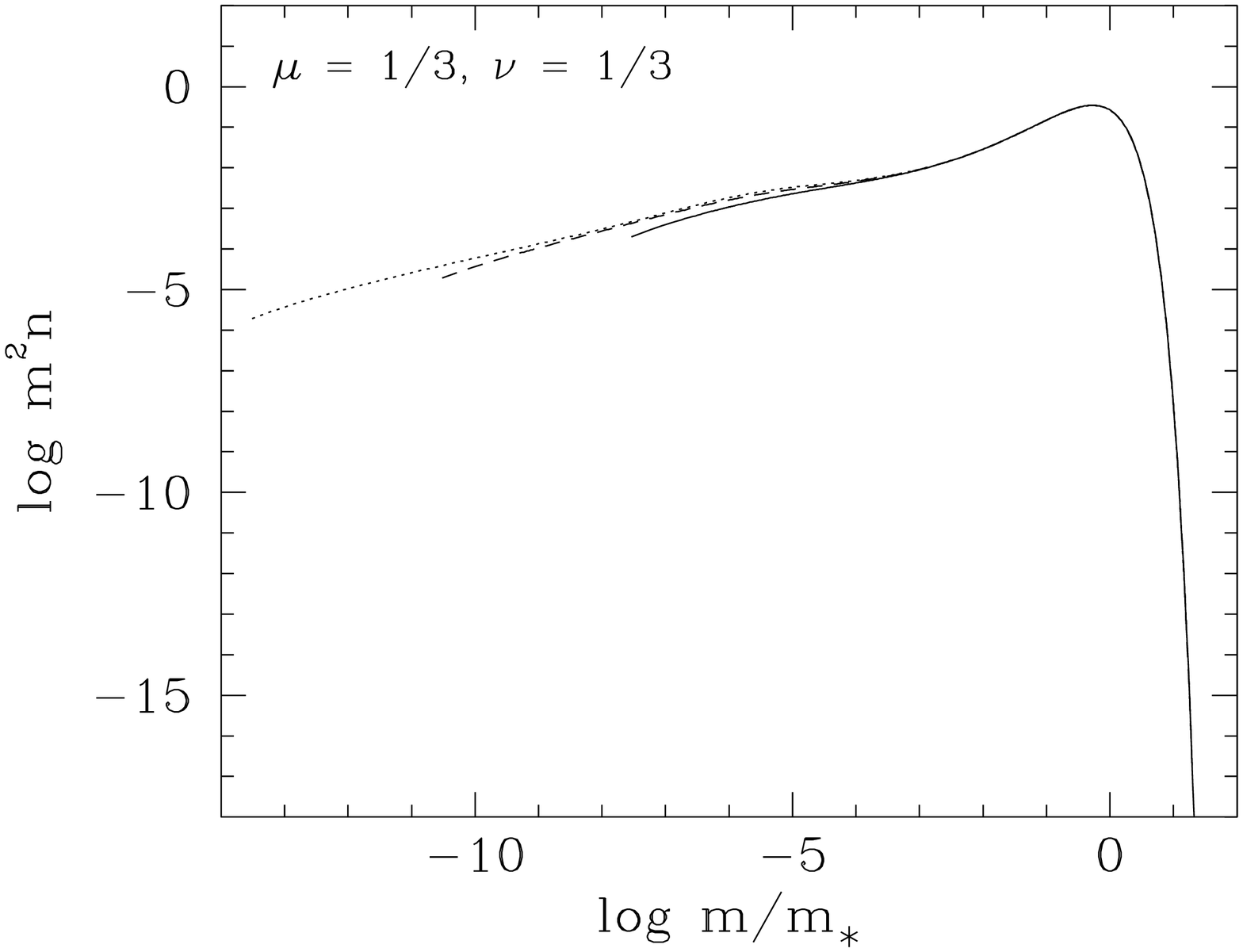}

{\bf Figure 4.}\ \ Approach to self-similar solution as $t \to \infty$ for
$(\mu,\nu) = (1/3,1/3)$.
The numerical solution is plotted for $t = 10^3$ (solid line), $10^4$ (dashed
line), and $10^5$ (dotted line) in the form of $\log m^2 n$ as a function
of $\log m/m_\ast$.

\vfill\eject

\epsfxsize=5truein \epsfbox[72 36 594 684]{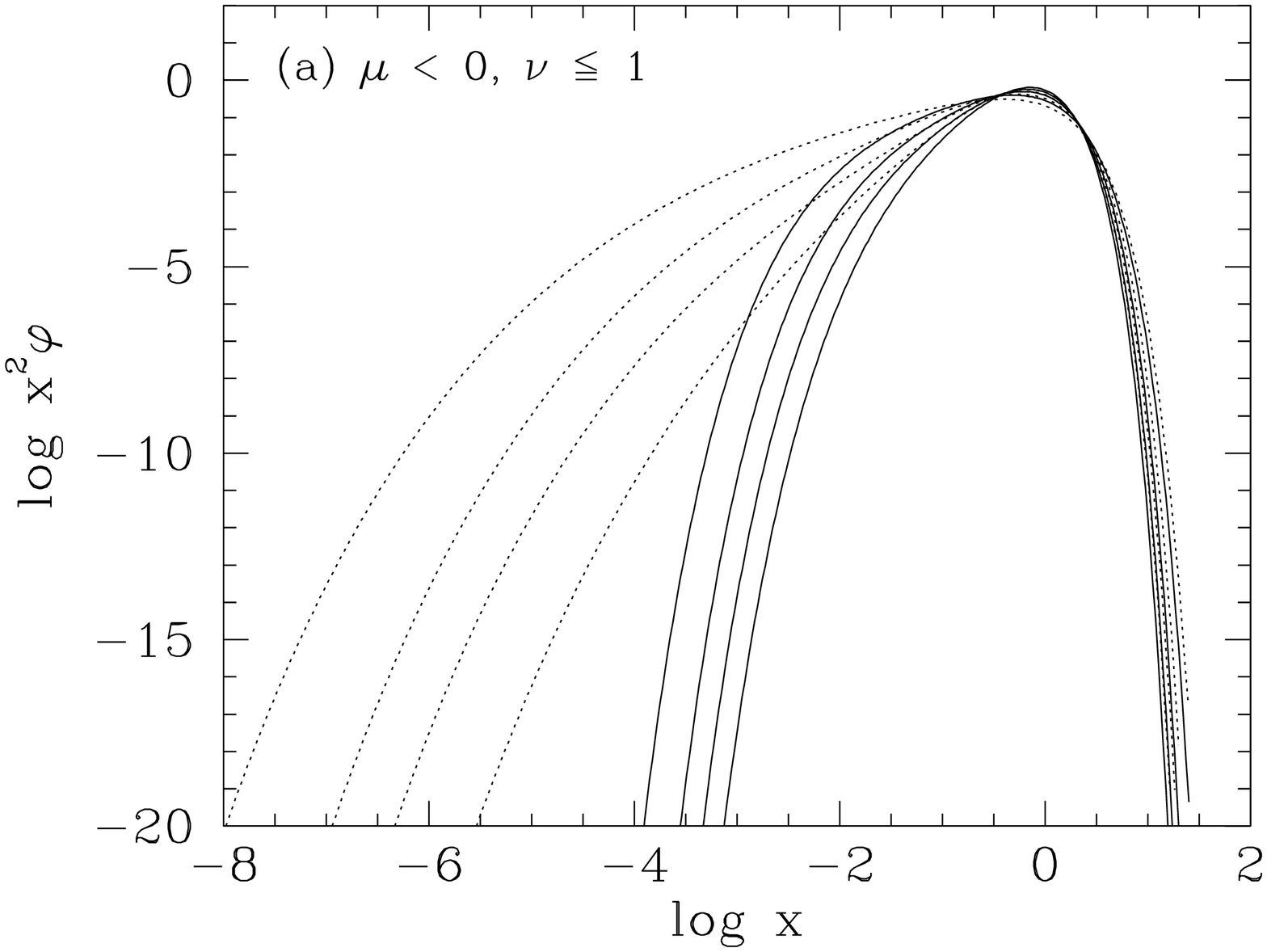}

\vfill\eject

\epsfxsize=5truein \epsfbox[72 36 594 684]{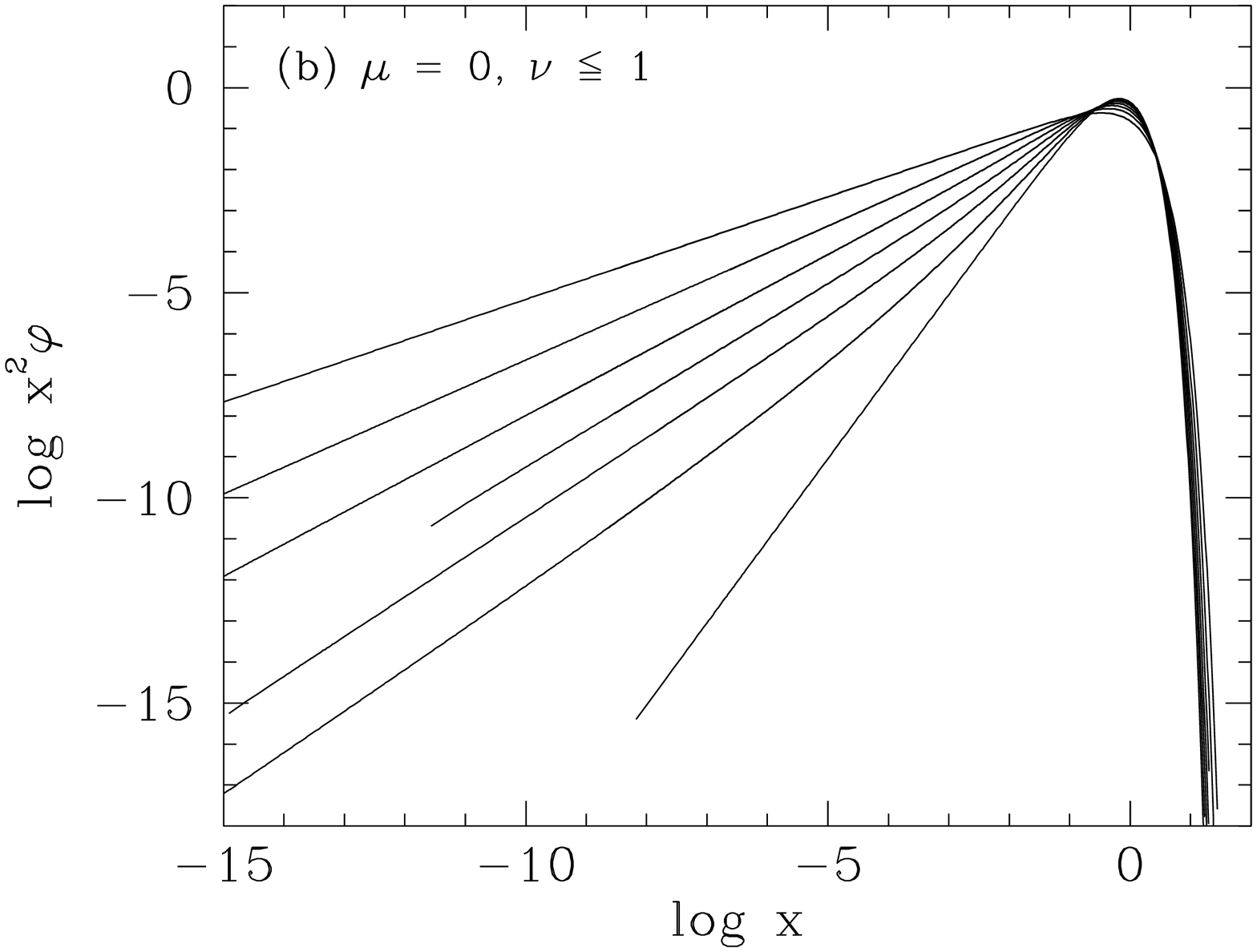}

\vfill\eject

\epsfxsize=5truein \epsfbox[72 36 594 684]{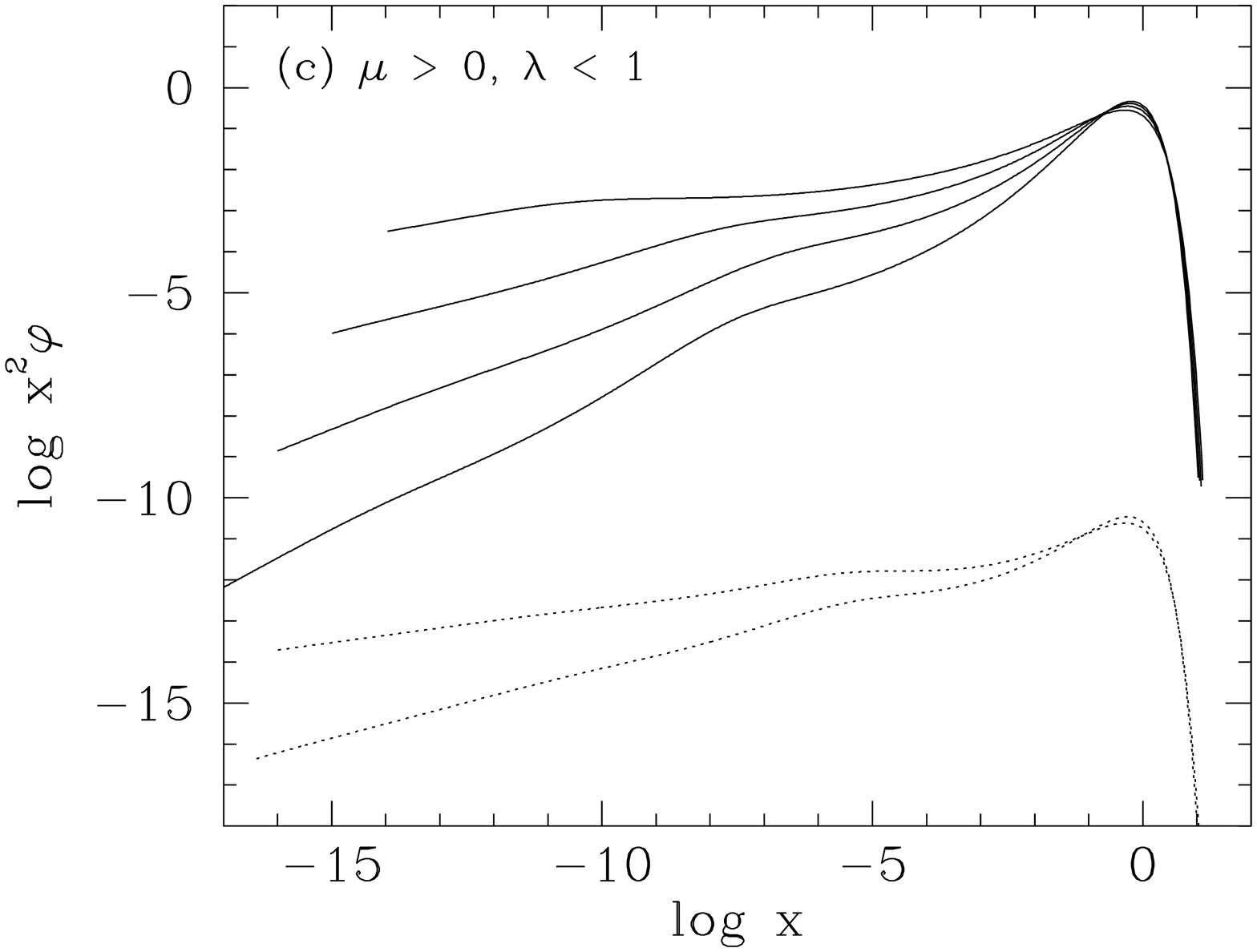}

{\bf Figure 5.}\ \ (a) Scaling function $\varphi(x)$ for the cases with
$\mu < 0$ and $\nu \le 1$.
The solid (dotted) lines in increasing width are for $\mu = -1/2$ ($-1/6$)
and $\nu = 0, 1/3, 2/3, 1$.
(b) $\varphi(x)$ for the cases with $\mu = 0$ and $\nu \le 1$.
The solid lines in increasing width are for $\nu = 0, 1/6,\ldots, 1$.
(c) $\varphi(x)$ for the cases with $\mu > 0$ and $\lambda < 1$.
The solid lines in increasing width are for $\mu = 1/6$ and $\nu = 1/6,
1/3, 1/2, 2/3$.
The dotted lines, offset vertically by $-10$, are for $\mu = 1/3$ and
$\nu = 1/3$ and $1/2$.

\vfill\eject

\epsfxsize=5truein \epsfbox[72 36 594 684]{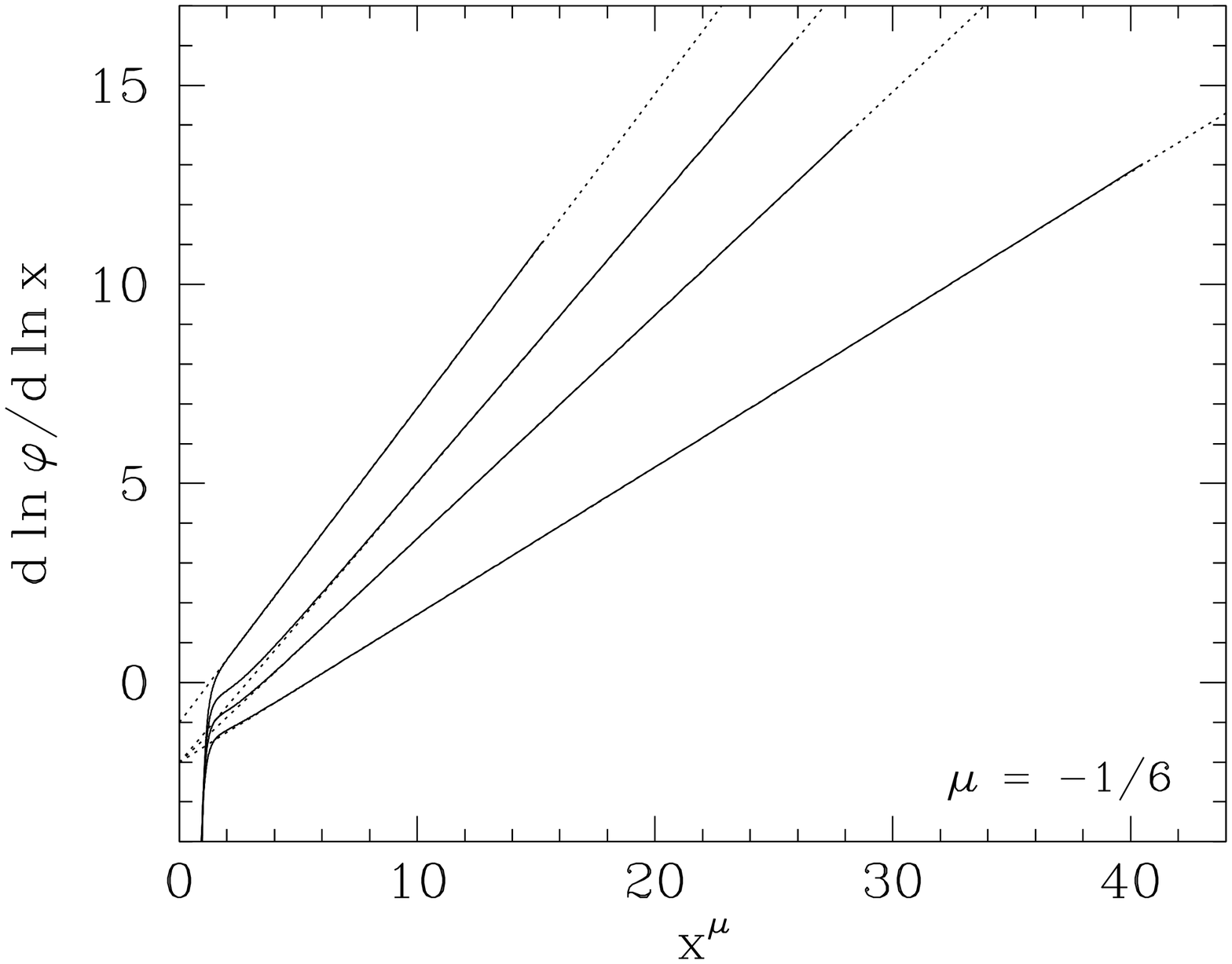}

{\bf Figure 6.}\ \ Logarithmic slope $\dlnpdlnx$ as a function of $x^\mu$
for the cases with $\mu = -1/6$ and $\nu = 0, 1/3, 2/3, 1$ (solid lines
from top to bottom).
The dotted lines are the asymptotes approached by the numerical results at
large $x^\mu$ (or small $x$).

\vfill\eject

\epsfxsize=5truein \epsfbox[72 36 594 684]{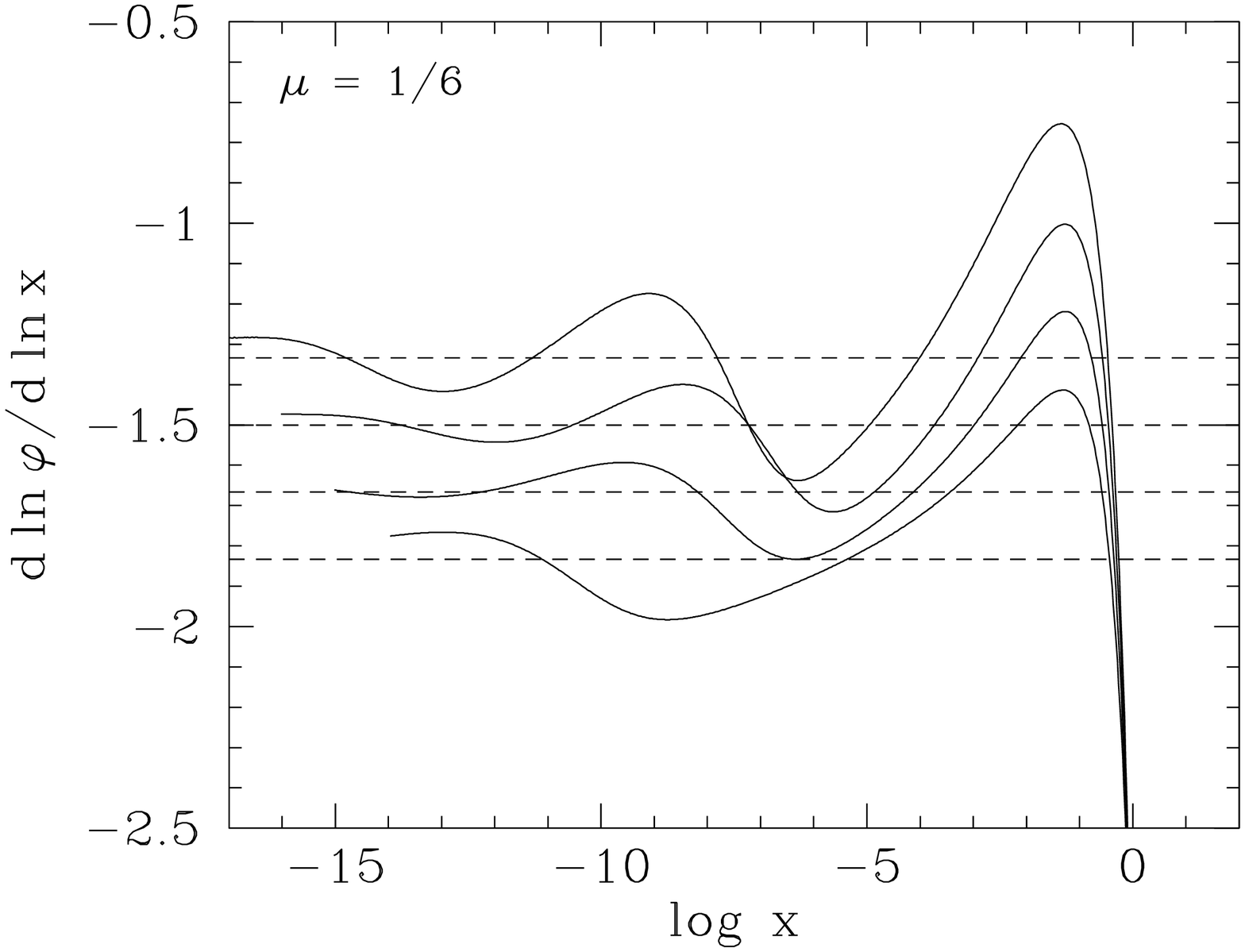}

{\bf Figure 7.}\ \ Logarithmic slope $\dlnpdlnx$ for the cases with $\mu =
1/6$ and $\nu = 1/6, 1/3, 1/2, 2/3$ (solid lines from top to bottom).
The dashed lines show the leading small-$x$ behavior predicted by
self-similar analysis: $\dlnpdlnx = -(1+\lambda)$.

\vfill\eject

\epsfxsize=5.5truein \epsfbox[18 144 576 702]{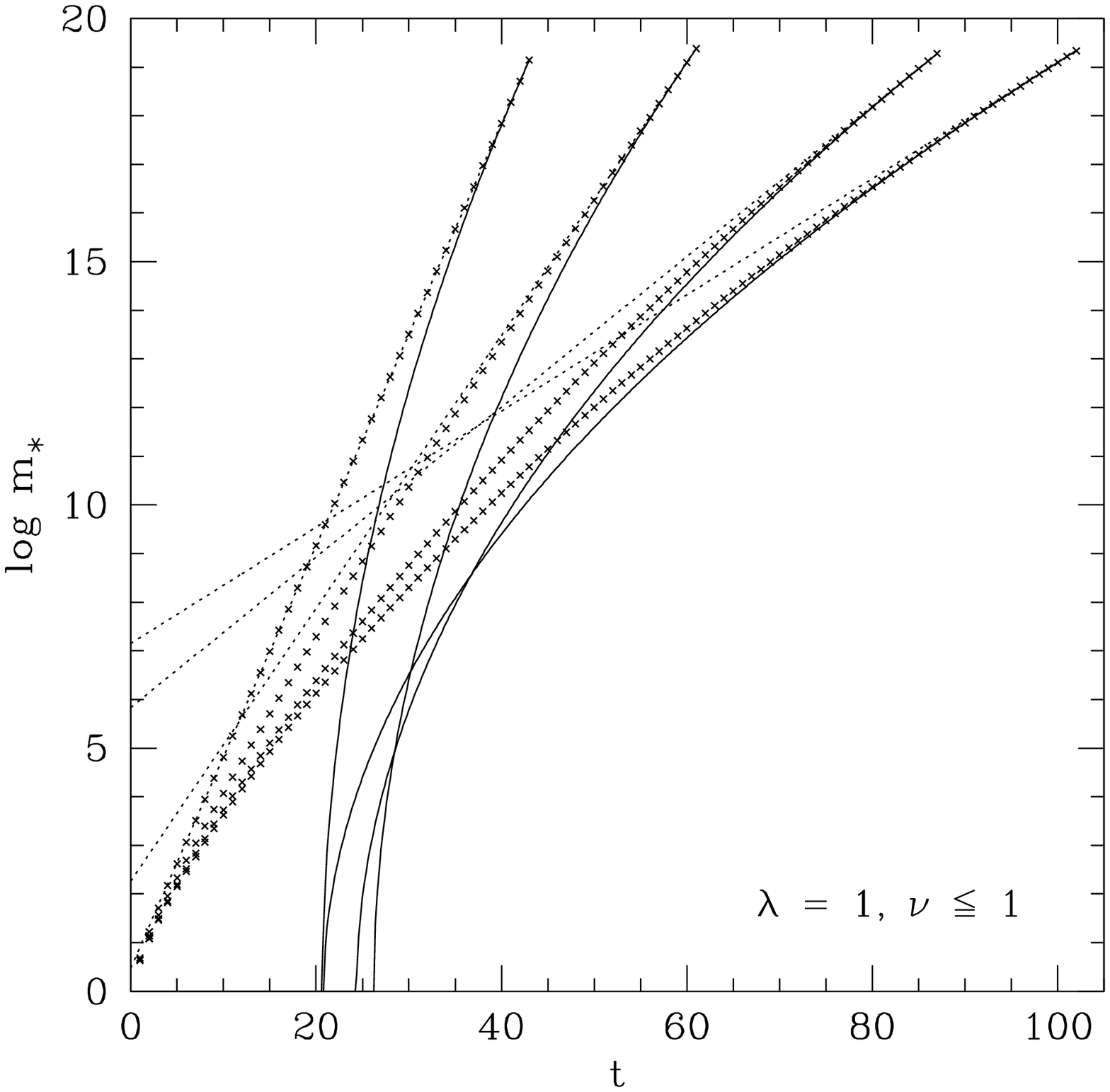}

{\bf Figure 8.}\ \ Characteristic mass $m_\ast(t)$ for the cases with
$\lambda = 1$ and $\nu \le 1$.
The points, from left to right, are the numerical results for $\mu = 0,
1/6, 1/3$, and $1/2$.
The dotted and solid lines show $\ln m_\ast = a + b t$ and $(\ln m_\ast)^2
= a + b t$ fitted to numerical results at the last two output times.

\vfill\eject

\epsfxsize=5truein \epsfbox[72 36 594 684]{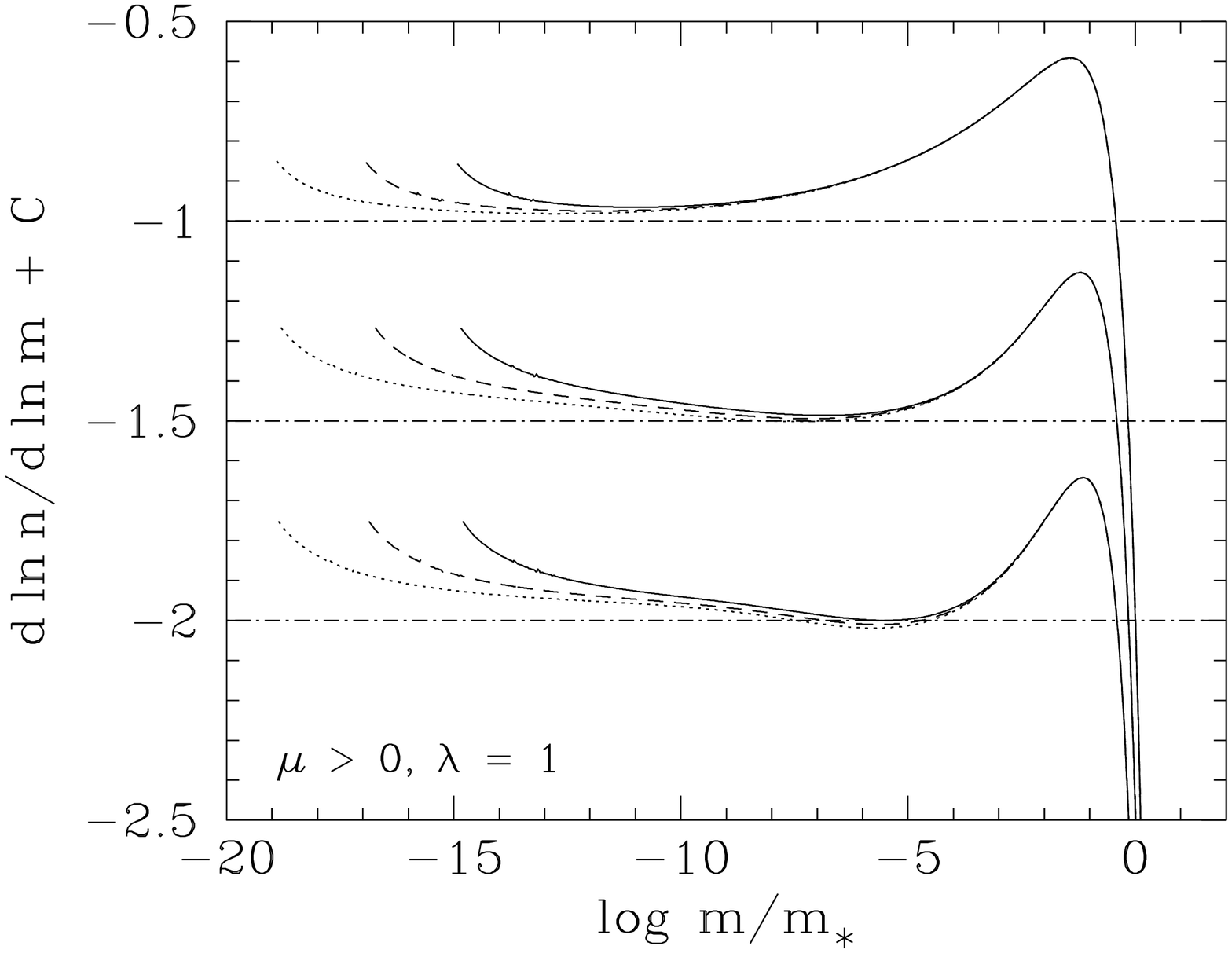}

{\bf Figure 9.}\ \ Logarithmic slope $\dlnndlnm$ for the cases with
$\mu > 0$ and $\lambda = 1$.
The top lines are the numerical results for $(\mu,\nu) = (1/6,5/6)$ at
$t = 47$ (solid line), $54$ (dashed line), and $61$ (dotted line), offset
vertically by $C = 1$.
The middle lines are for $(\mu,\nu) = (1/3,2/3)$ at $t = 63$, $74$, and
$87$, with $C = 0.5$, and the bottom lines are for $(\mu,\nu) = (1/2,1/2)$
at $t = 71$, $86$, and $102$, with $C = 0$.

\vfill\eject

\epsfxsize=5truein \epsfbox[72 36 594 684]{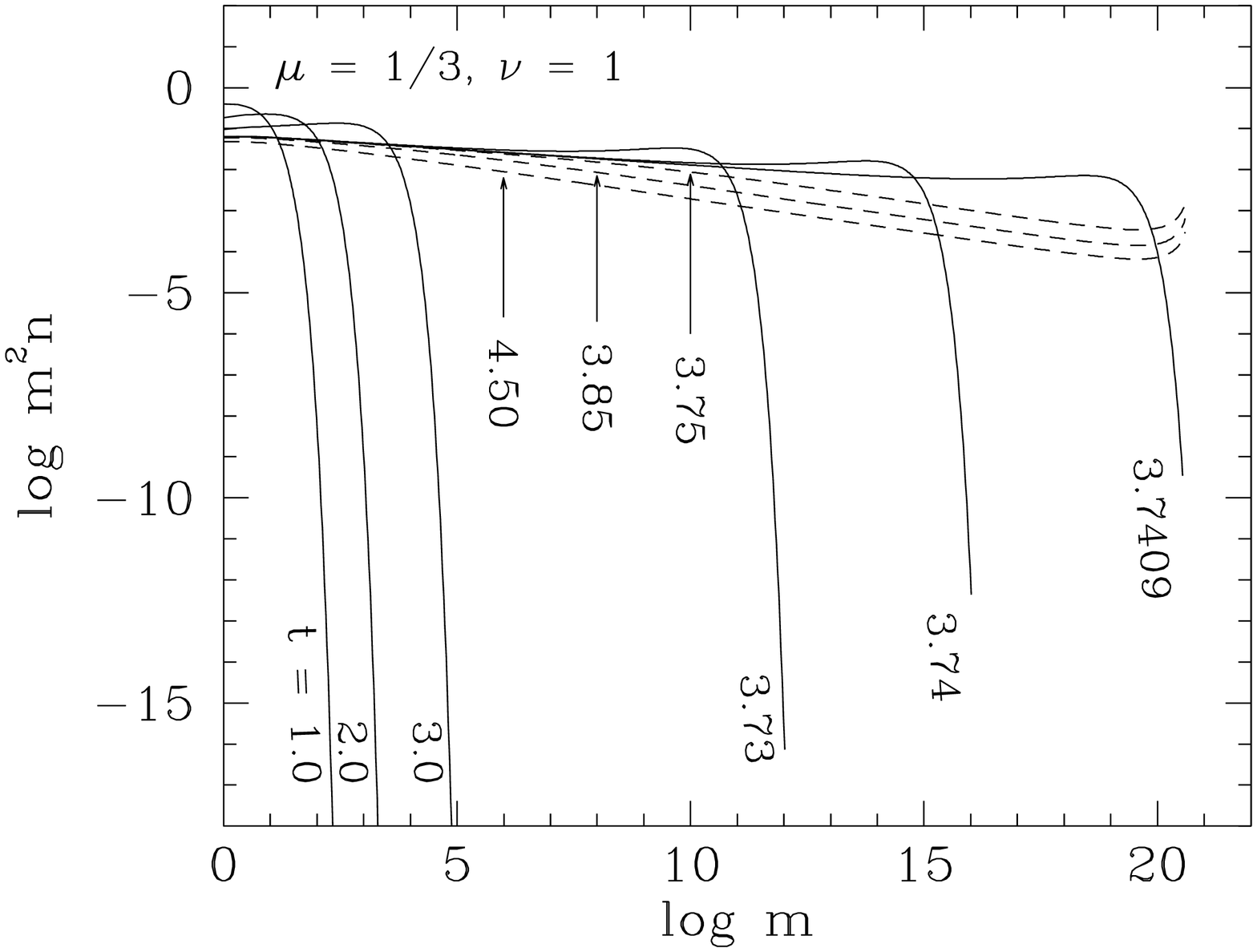}

{\bf Figure 10.}\ \ Evolution of the mass spectrum for $(\mu,\nu) = (1/3,1)$
at $t \le t_0$ (solid lines) and $t > t_0$ (dashed lines).

\vfill\eject

\epsfxsize=5truein \epsfbox[72 36 594 684]{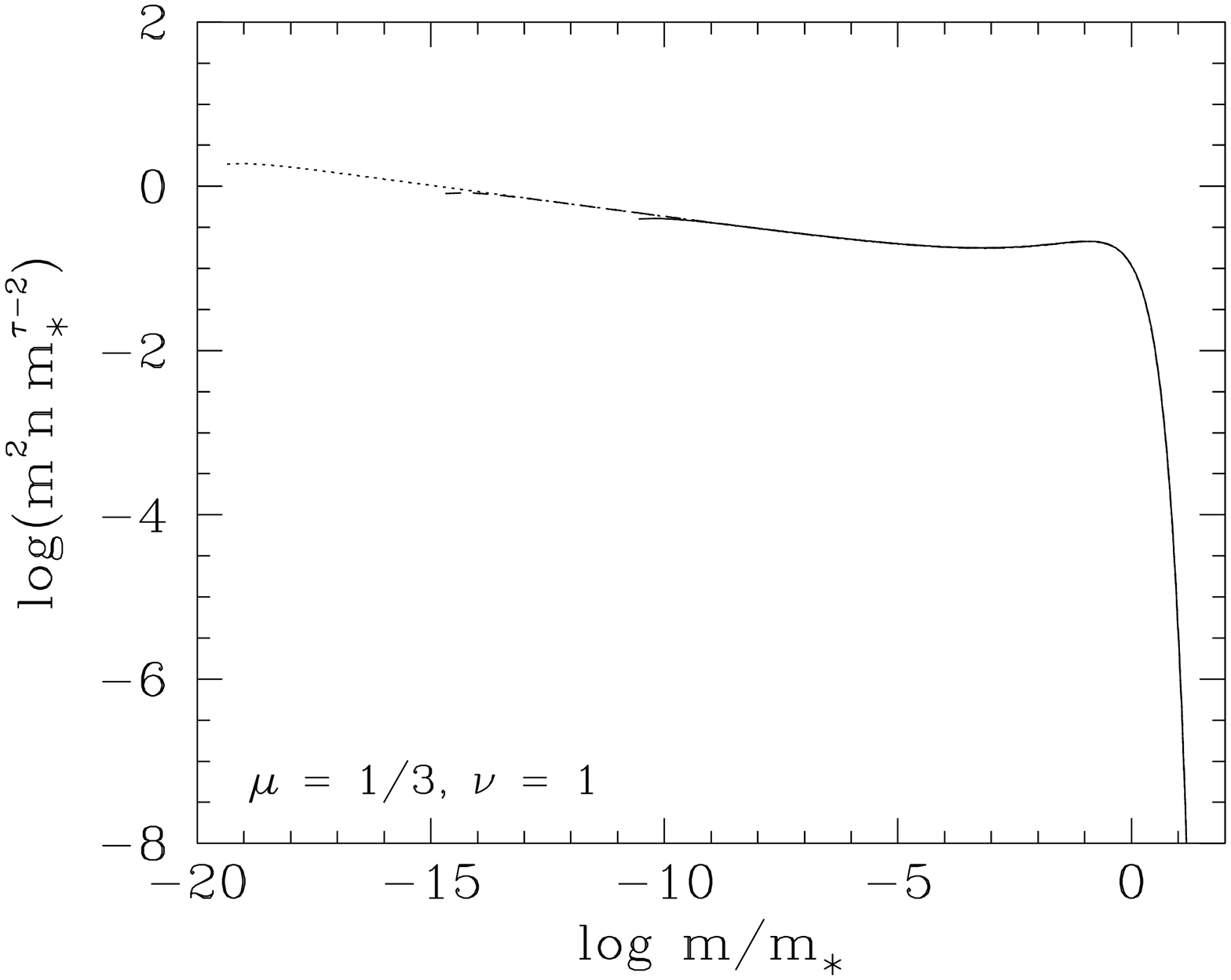}

{\bf Figure 11.}\ \ Approach to self-similar solution as $t \uparrow t_0$
for $(\mu,\nu) = (1/3,1)$.
The numerical solution is plotted for $t = 3.73$ (solid line), $3.74$
(dashed line), and $3.7409$ (dotted line) in the form of
$\log(m^2 n\, m_\ast^{\tau-2})$ as a function of $\log m/m_\ast$.

\vfill\eject

\epsfxsize=5truein \epsfbox[72 36 594 684]{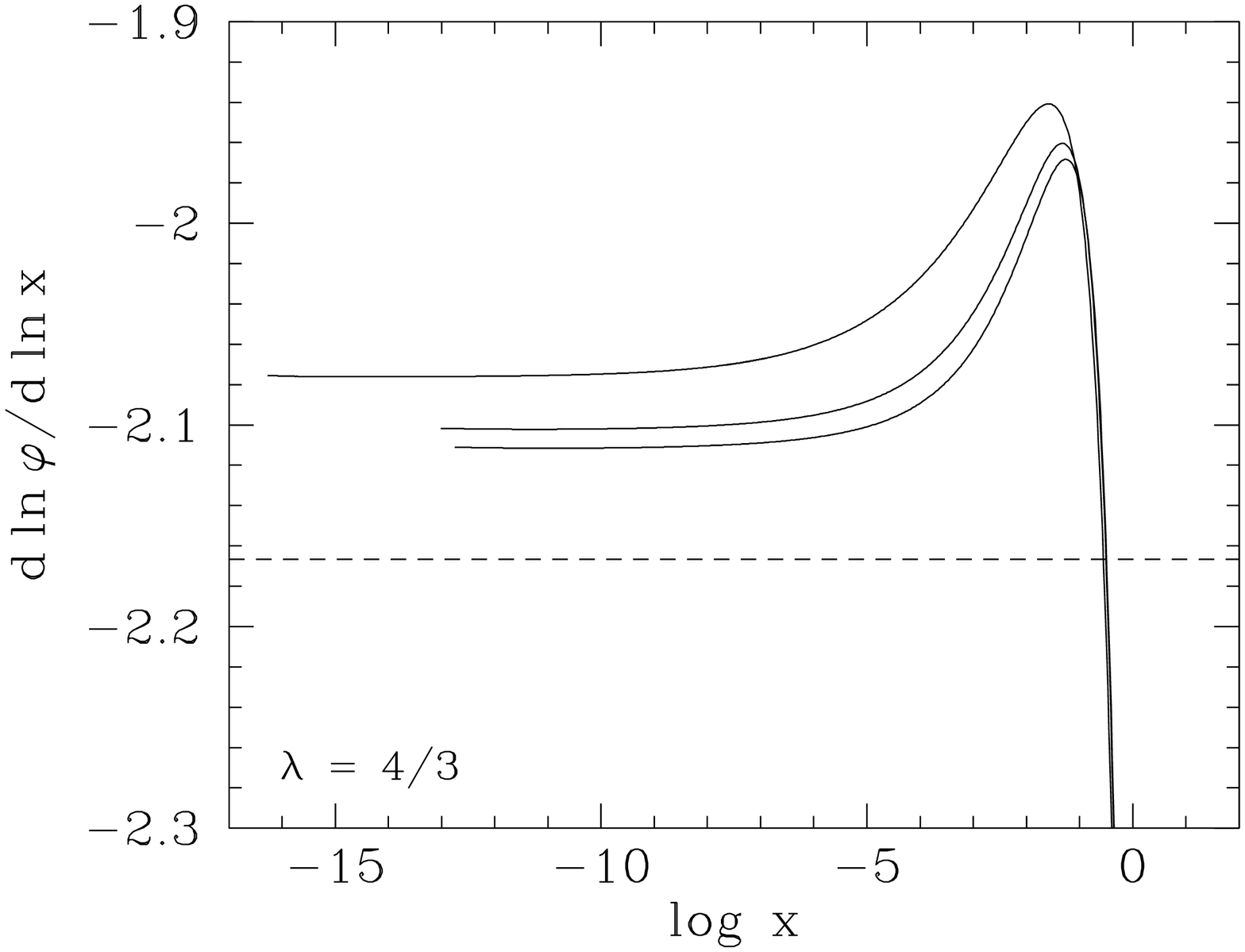}

{\bf Figure 12.}\ \ Logarithmic slope $\dlnpdlnx$ for the cases with 
$\lambda = 4/3$.
The solid lines from top to bottom are the numerical results for
$(\mu,\nu) = (1/3,1)$, $(1/2,5/6)$, and $(2/3,2/3)$, and
the dashed line is the analytic prediction for the small-$x$ behavior:
$\dlnpdlnx = -\tau = -(\lambda + 3)/2 = -13/6$.

\vfill\eject

\epsfxsize=5.5truein \epsfbox[18 144 576 702]{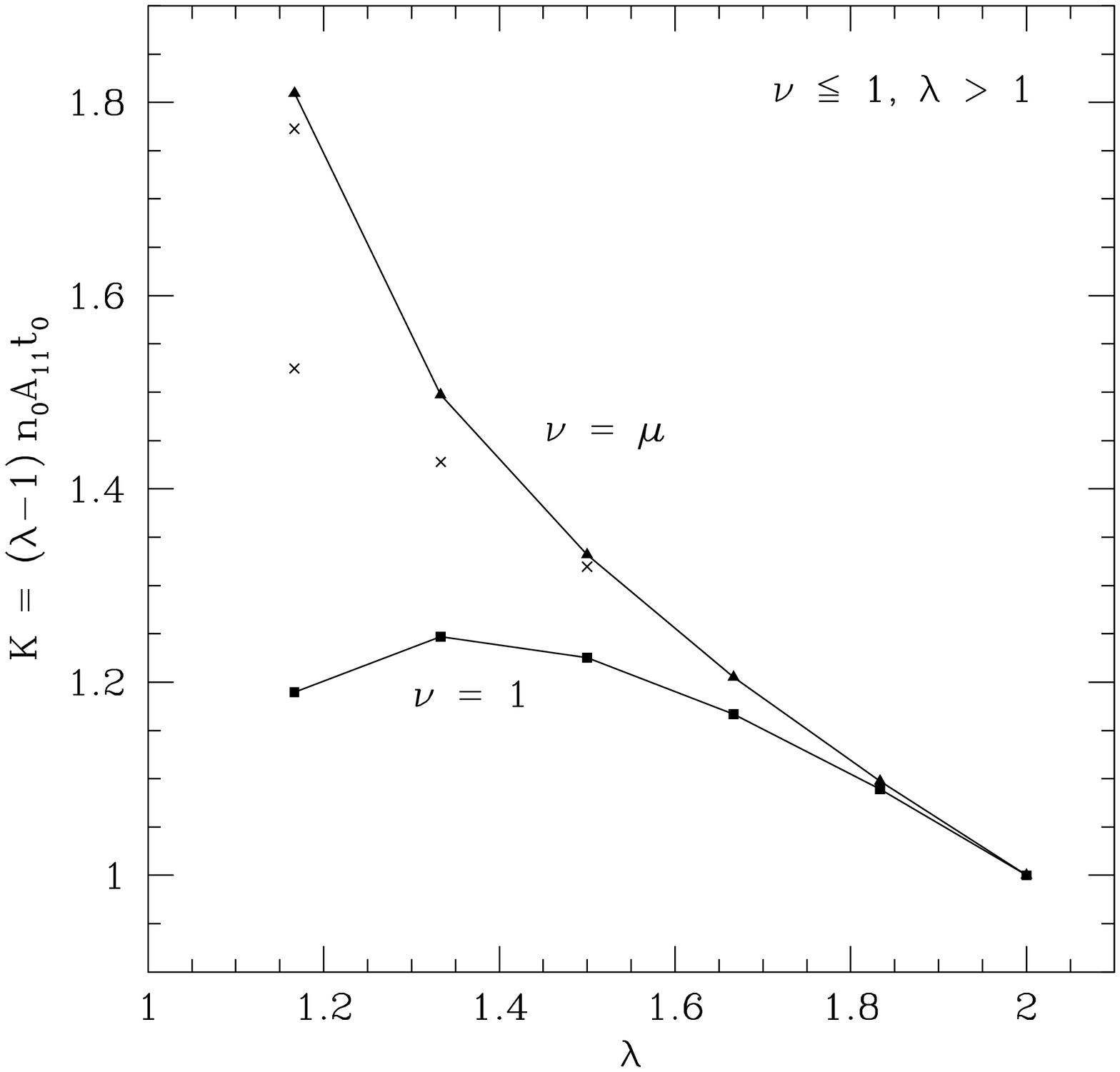}

{\bf Figure 13.}\ \ Parameter $K = (\lambda - 1) n_0 A_{11} t_0$ for the
cases with $\nu \le 1$ and $\lambda > 1$ listed in table 1.
The cases with $\nu = 1$ and $\nu = \mu$ are represented by squares and
triangles, respectively, and the remaining cases are represented by crosses.

\vfill\eject

\epsfxsize=5.5truein \epsfbox[72 36 594 684]{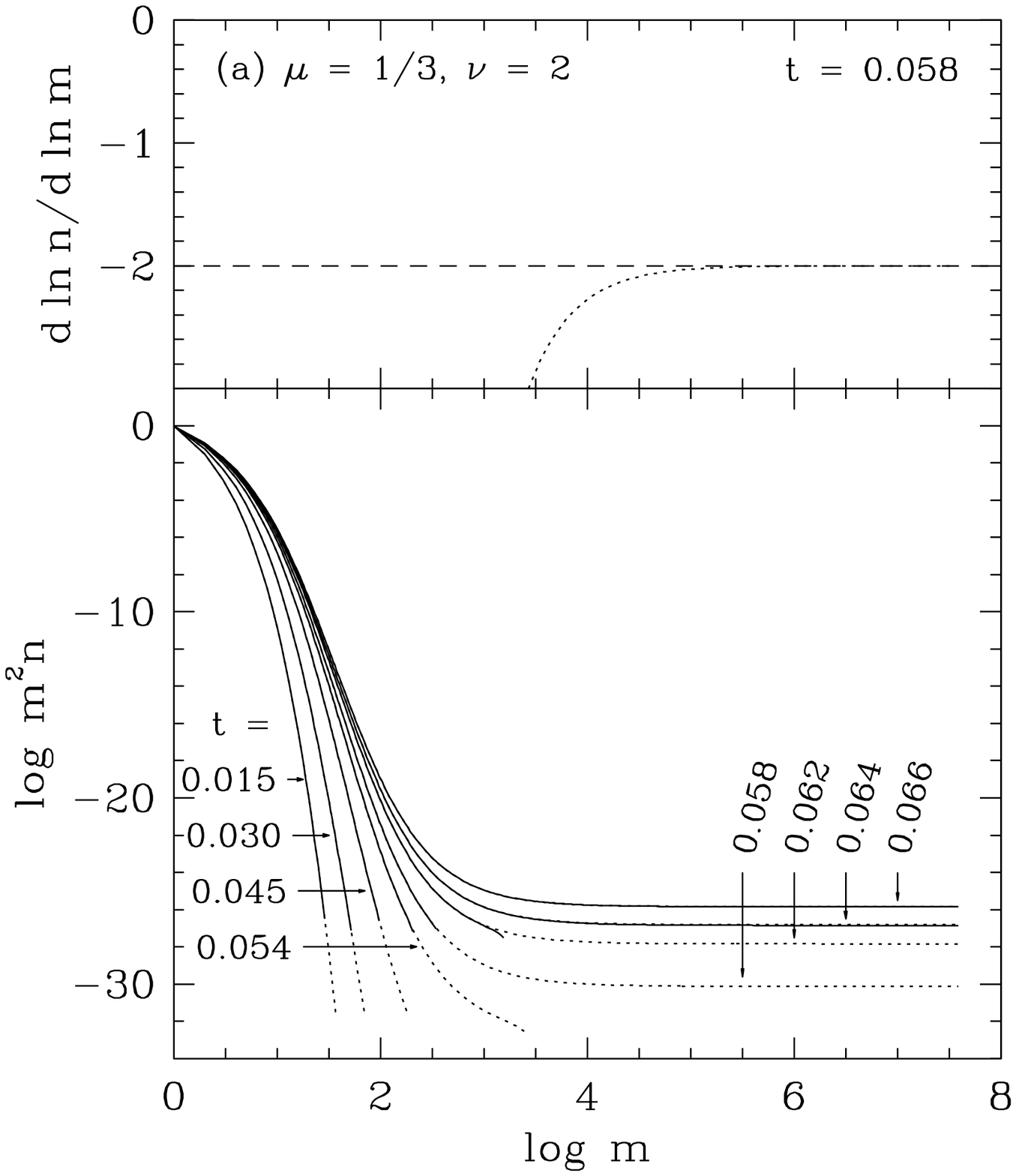}

\vfill\eject

\epsfxsize=5.5truein \epsfbox[72 36 594 684]{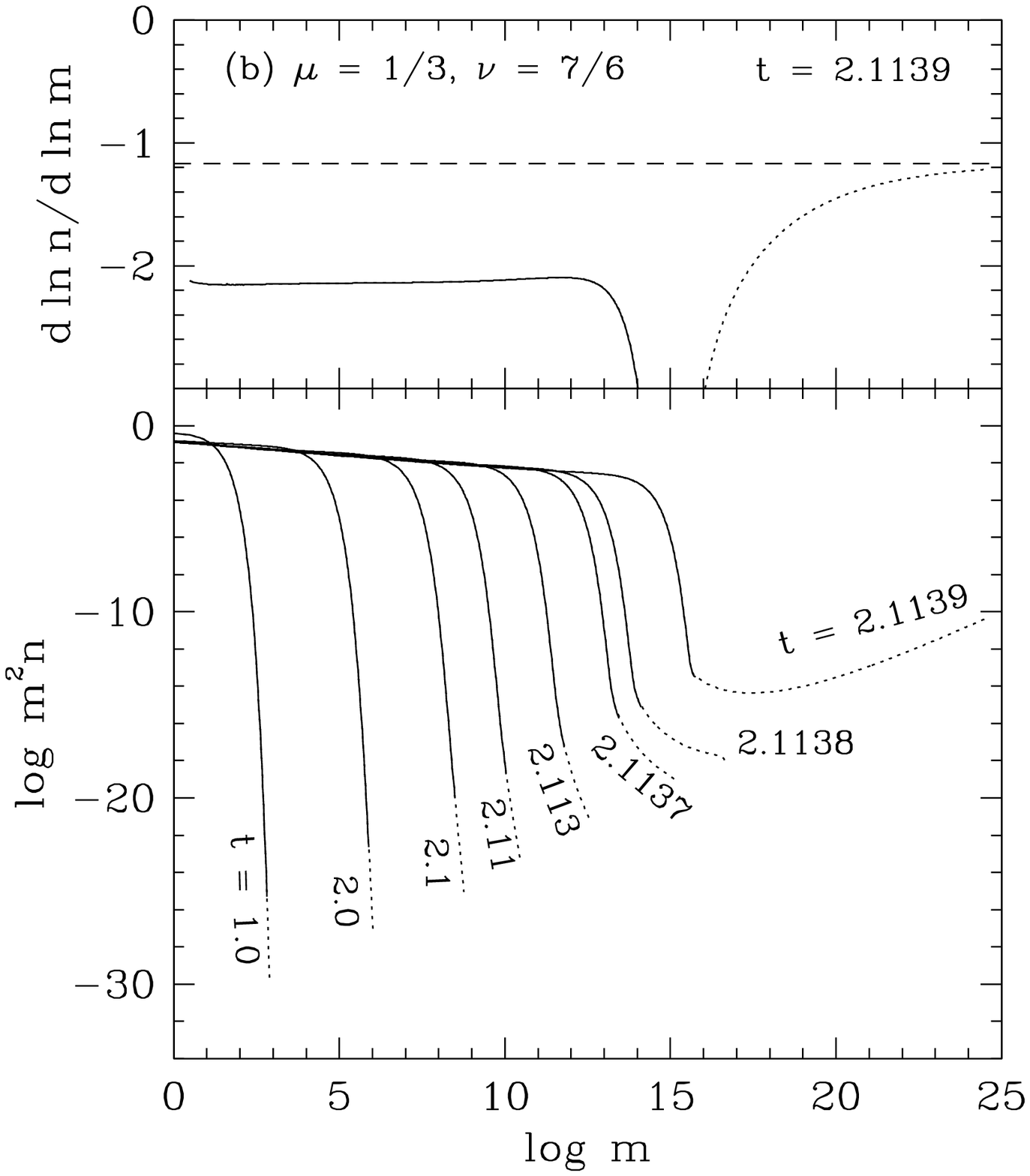}

{\bf Figure 14.}\ \ Evolution of the mass spectrum ({\it lower panels}) and
logarithmic slope $\dlnndlnm$ at the specified time ({\it upper panels})
for $\mu = 1/3$ and $\nu$ equal to (a) $2$ and (b) $7/6$.
The numerical solutions with $\nmin = 10^{-30}$ (solid lines) and $10^{-35}$
(dotted lines) are shown.
In the upper panels, the dashed lines indicate $\dlnndlnm = -\nu$.

\vfill\eject
\end